\documentclass{aa}
\usepackage{graphicx,fixltx2e}
\usepackage{txfonts}
\usepackage{color}
\usepackage{natbib} 
\bibpunct{(}{)}{;}{a}{}{,}

\begin{document}
 \title{The energy of waves in the photosphere and lower
   chromosphere: II.~Intensity statistics}

   \author{C. Beck\inst{1,2} \and R. Rezaei\inst{3} \and K.G. Puschmann\inst{4}}
        
   \titlerunning{The energy of waves: II.~Intensity statistics}
  \authorrunning{C. Beck, R. Rezaei, K.G. Puschmann}  
   \institute{Instituto de Astrof\'isica de Canarias
   \and Kiepenheuer-Institut f\"ur Sonnenphysik
   \and Leibniz-Institut f\"ur Astrophysik
       }
 \date{Received xxx; accepted xxx}

\abstract{The energy source that maintains the solar chromosphere is still
  undetermined, but leaves its traces in observed intensities.}{We investigate
  the statistics of the intensity distributions as function of the wavelength for \ion{Ca}{ii}\,H and the \ion{Ca}{ii}\,IR line at 854.2\,nm to estimate the energy content in the observed intensity fluctuations.}{We derived the intensity variations at different heights of the solar atmosphere as given by the line wings and line cores of the two spectral lines. We converted the observed intensities to absolute energy units employing reference profiles calculated in non-local thermal equilibrium (NLTE). We also converted the observed intensity fluctuations to corresponding brightness temperatures assuming LTE.}{The root-mean-square (rms) fluctuations of the emitted intensity are about 0.6 (1.2) Wm$^{-2}$ster$^{-1}$pm$^{-1}$ near the core of the \ion{Ca}{ii}\,IR line at 854.2\,nm (\ion{Ca}{ii}\,H), corresponding to relative intensity fluctuations of about 20\,\% (30\,\%). Maximum fluctuations can be up to 400\,\%. For the line wing, we find rms values of about 0.3 Wm$^{-2}$ster$^{-1}$pm$^{-1}$  for both lines, corresponding to relative fluctuations below 5\,\%. The relative rms values show a local minimum for wavelengths forming at about 130\,km height, but otherwise increase smoothly from the wing to the core, i.e., from photosphere to chromosphere. The corresponding rms brightness temperature fluctuations are below 100\,K for the photosphere and up to 500\,K in the chromosphere. The skewness of the intensity distributions is close to zero in the outer line wing and positive throughout the rest of the line spectrum, caused by a frequent occurrence of high-intensity events. The skewness shows a pronounced local maximum on locations with photospheric magnetic fields for wavelengths in between the line wing and the line core ($z\approx\,150-300$\,km), and a global maximum at the very core ($z\approx\,1000$\,km) for both magnetic and field-free locations.}
{The energy content of the intensity fluctuations is insufficient to create a similar temperature rise in the chromosphere as predicted in most reference models of the solar atmosphere. The increase of the rms fluctuations with height indicates the presence of upwards propagating acoustic waves with an increasing oscillation amplitude. The intensity and temperature variations show a clear increase of the dynamics from photosphere towards the chromosphere, but fall short of fully dynamical chromospheric models by a factor of about five. The enhanced skewness between photosphere and lower solar chromosphere on magnetic locations indicates a mechanism which solely acts on magnetized plasma. Possible candidates are the Wilson depression, wave absorption, or magnetic reconnection.}
\maketitle
\section{Introduction}
The solar chromosphere shows prominent emission lines when viewed near the
solar limb, e.g., during an eclipse, but also on the solar disc chromospheric spectral lines such as H$\alpha$, \ion{Ca}{ii}\,H and K, \ion{Mg}{ii}\,h and k, or the \ion{Ca}{ii}\,IR triplet revert to transient emission. Because the gas density is so low that the material is transparent and the radiation temperature is decreasing above the continuum forming layers in the photosphere, the emission lines require an energy supply other than radiative transfer. The different types of possible heating mechanisms vary from purely mechanical processes to all processes that can be related to the presence of magnetic field lines \citep[see, e.g.,][]{biermann_48,schatzman1949,liu1974,anderson+athay1989,davila+chitre1991,rammacher+ulmschneider1992,narain+ulmschneider1996,carlsson+etal2007,rezaei+etal2007,beck+etal2008,fontenla+etal2008,beck+etal2009,khomenko+collados2012}.

In addition to the fact that ``the'' heating mechanism of the chromosphere
could not be identified yet, the exact amount of energy required to maintain
the chromosphere as it is observed is not fully clear. There exists a series of
static atmospheric stratification models for the solar photosphere and
chromosphere
\citep[e.g.,][]{gingerich+etal1971,vernazza+etal1976,vernazza+etal1981,fontenla+etal2006,avrett+loeser2008}
that were determined from temporally and/or spatially averaged spectra. These
models mainly share the existence of a temperature reversal at a certain
height in the solar atmosphere, the location of the temperature minimum. Using
temporally and spatially resolved spectra, it seems that the assumption of a
static background temperature with minor variation around it is not fulfilled
\citep[e.g.,][]{liu+skumanich1974,kalkofen+etal1999,carlsson+stein1997,rezaei+etal2008}.
The spectra indicate in some cases that no chromospheric temperature rise is
present at all \citep{liu+smith1972,cram+dame1983,rezaei+etal2008}, as is also
required by the observations of CO molecular lines in the lower chromosphere
\citep[][]{ayres+testerman1981,ayres2002} that only form at temperatures below
about 4000\,K. Numerical hydrodynamical (HD) simulations of the chromosphere show even lower temperatures down to 2000\,K \citep[][]{wedemeyer+etal2004,leenarts+etal2011}.

In the first paper of this series \citep[][BE09]{beck+etal2009} we
investigated the energy content of velocity oscillations in several
photospheric spectral lines and the chromospheric \ion{Ca}{ii}\,H line. The
main finding of BE09 was that the energy of the root-mean-square (rms)
velocity of a spectral line in the wing of \ion{Ca}{ii}\,H, whose formation
height was estimated to be about 600\,km, was already below the chromospheric
energy requirement of 4.3\,kWm$^{-2}$ given by \citet{vernazza+etal1976}, with a
further decrease of the energy content towards higher layers. The observations used had a spatial resolution of about 1$^{\prime\prime}$, similar to those of
\citet{fossum+carlsson2005,fossum+carlsson2006} that also found an insufficient
energy in intensity oscillations observed with TRACE. \citet{wedemeyer+etal2007} and \citet{kalkofen2007} demonstrated later that the spatial resolution can be critical for the determination of the energy content, because the spatial smearing hides high-frequency oscillations with their corresponding short wavelengths. 

Recently, several papers addressed the energy flux of acoustic and gravity waves using data of higher spatial resolution from, e.g., the G\"ottingen Fabry-Perot Interferometer \citep[GFPI,][]{puschmann+etal2006}, the Interferometric BI-dimensional Spectrometer \citep[IBIS,][]{cavallini2006}, or the Imaging Magnetograph eXperiment (IMaX) onboard of the Sunrise balloon mission \citep{jochum+etal2003,gandorfer+etal2006,martinezpillet+etal2011}. \citet{straus+etal2008} found an energy flux of solar gravity waves of 5\,kWm$^{-2}$ at the base of the chromosphere in simulations and observations, usually an order of magnitude higher than the energy flux of acoustic waves at the same height. \citet{nazi+etal2009,nazi+etal2010} found an acoustic energy flux of 3 and $>$\,6\,kWm$^{-2}$ at a height of 250\,km, using GFPI and IMaX data, respectively \citep[see also Table 2 in][]{nazi+etal2010}. Much of the wave energy was found in the frequency range above the acoustic cut-off frequency of about 5\,mHz; these high-frequency waves are able to propagate into the chromosphere without strong (radiative) losses \citep[cf.~also][]{carlsson+stein2002,reardon+etal2008}. The energy content was, however, basically determined from {\em photospheric} spectral lines, and thus refers to the upper photosphere, where also at lower spatial resolution sufficient mechanical energy is seen (BE09). Interestingly, the (lower) chromospheric lines used in \citet{straus+etal2008}, \ion{Na}{i} D and \ion{Mg}{i} $b_2$, showed an energy flux {\em below} 4\,kWm$^{-2}$ (their Fig.~3). It is therefore unclear how reliable the extrapolation of the photospheric energy fluxes actually is. 

Another point to be considered is the influence of the photospheric magnetic
fields on the chromospheric behaviour \citep[e.g.,][]{cauzzi+etal2008}. Photospheric magnetic fields also cause an enhancement of the chromospheric emission, even if the underlying process may be a kind of wave propagation in the end
\citep[e.g.,][]{hasan2000,rezaei+etal2007,beck+etal2008,hasan+etal2008,khomenko+etal2008,vigeesh+etal2009,fedun+etal2011}. The usage of velocity oscillations to estimate the energy content is in general also less direct than to use the intensity spectrum. If energy is deposited into the upper solar atmosphere by a process without a clear observable signature, either because of missing spatial or temporal resolution, or because of physical reasons such as for the case of transversal wave modes, the emitted intensity will still react to the energy deposit, even if the non-local thermal equilibrium (NLTE) conditions in the chromosphere make the reaction neither instantaneous nor necessarily linear to the amount of deposited energy. For instance, \citet{reardon+etal2008} found evidence that the passage of shock fronts induces turbulent motions at chromospheric heights. The dissipation of the energy of these turbulent motions then would effect an energy transfer from mechanical to radiative energy, but with some time delay and over some period longer than the duration of the shock passage itself \citep[see also][]{verdini+etal2012}.

In this contribution, we investigate the intensity statistics of two chromospheric spectral lines, \ion{Ca}{ii}\,H and the \ion{Ca}{ii}\,IR line at 854.2\,nm, to estimate the energy contained in their intensity variations at all wavelengths from the wing to the very line core. Section \ref{sec_obs} describes the observations used. The data reduction steps are detailed in Sect.~\ref{add_data}. In Sect.~\ref{sec_int}, we investigate the intensity statistics of the two
lines. The results are summarized and discussed in Sect.~\ref{summ_disc},
whereas Sect.~\ref{concl} provides our conclusions.
\begin{figure*}
\centering
\fbox{\begin{minipage}{8.8cm}\centerline{Observation No.~3}$ $\\\resizebox{8.8cm}{!}{\hspace*{.7cm}\includegraphics{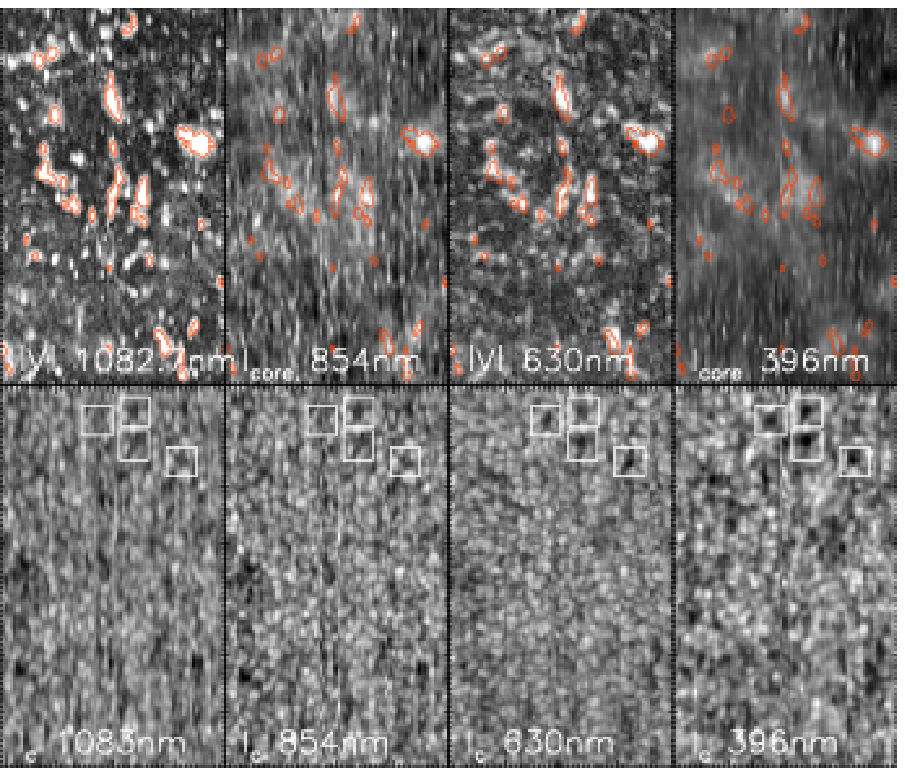}}$ $\\$ $\\\end{minipage}}\hspace*{.25cm}\fbox{\begin{minipage}{8.3cm}\centerline{Observation No.~4}$ $\\\resizebox{8.3cm}{!}{\hspace*{.7cm}\includegraphics{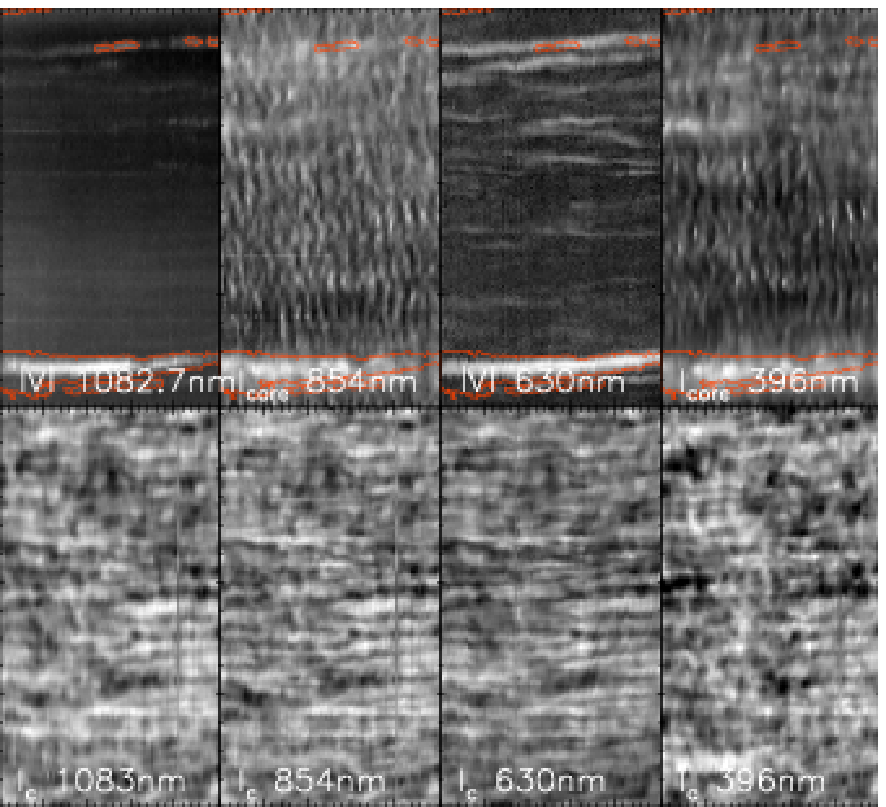}}$ $\\$ $\\\end{minipage}}
\caption{Overview maps of the observation No.~3 ({\em left panel}) and No.~4 ({\em right panel}). {\em Bottom row, left to right}: continuum intensity at 1083\,nm, 854\,nm, 630\,nm, and 396\,nm. {\em Top row, left to right}: unsigned wavelength-integrated Stokes $V$ signal around \ion{Si}{i} at 1082.7\,nm, line-core intensity of the \ion{Ca}{ii}\,IR line at 854.2\,nm, unsigned wavelength-integrated Stokes $V$ signal around 630\,nm, line-core intensity of \ion{Ca}{ii}\,H at 396.85\,nm. The {\em red contours} outline strong polarisation signals. The {\em white squares} denote co-spatial local darkenings.\label{overview_2009}}
\end{figure*}
\begin{table}
\caption{Overview of the observations.\label{tab_obs}}
\begin{tabular}{cccccc}
No. & date & type & $t_{\rm int}$ & size & \ion{Ca}{ii}\,IR\cr\hline
1 & 24/07/06 & map & 6.6 sec & 75$^{\prime\prime}\times 70^{\prime\prime}$ & no\cr
2 & 24/07/06 & time series & 3.3 sec & 60 min & no \cr
3 & 29/08/09 & map & $\sim 30$\, sec & 40$^{\prime\prime}\times 70^{\prime\prime}$ &yes \cr
4 & 08/09/09 & time series & 5 sec & 50 min &yes \cr
\end{tabular}
\end{table}
\section{Observations\label{sec_obs}}
We used four different observations, two large-area maps and two time series
of about one hour duration, all observed in quiet Sun (QS) on disc centre with real-time seeing correction by the Kiepenheuer-Institute Adaptive Optics System \citep[KAOS,][]{vdluehe+etal2003}. One of the large-area maps and one time series were taken in the morning of 24 July 2006. They are
already described in detail in \citet[][BE08]{beck+etal2008} and BE09; we thus only shortly summarize their characteristics. The large-area map covered an area of 75$^{\prime\prime} \times$ 70$^{\prime\prime}$, taken by moving the
0\farcs5 wide slit of the POlarimetric LIttrow Spectrograph
\citep[POLIS,][]{beck+etal2005b} at the German Vacuum Tower Telescope \citep[VTT,][]{schroeter+soltau+wiehr1985} in steps of 0\farcs5 across the solar
image. For the time series, the cadence of co-spatial positions was about 21\,seconds. POLIS provided the Stokes vector around 630\,nm together
with intensity profiles of \ion{Ca}{ii}\,H. The spatial sampling along the slit
was 0\farcs29 for \ion{Ca}{ii}\,H and half of that for 630\,nm.  The polarimetric
data at 630\,nm were calibrated with the methods described in
\citet{beck+etal2005b,beck+etal2005a}. These observations will be labeled
No.~1 (large-area map on 24 July 2006) and No.~2 (times series on 24 July 2006)
in the following; Table \ref{tab_obs} lists additionally the integration time
$t_{\rm int}$ per scan step. 

The third observation is another large-area map on disc centre from 28 August 2009, UT 08:33:42 until 09:42:08, this time taken simultaneously with POLIS, the Tenerife Infrared Polarimeter \citep[TIP,][]{martinez+etal1999,collados+etal2007}, and a PCO 4000 camera on the output port of the main spectrograph of the VTT to record spectra of the \ion{Ca}{ii}\,IR line at 854.2\,nm . TIP observed the wavelength range around 1083\,nm, including a photospheric \ion{Si}{i} line at 1082.7\,nm and the chromospheric \ion{He}{i} line at 1083\,nm. The slit width of the main spectrograph was 0\farcs36 and the spatial sampling along the slit was 0\farcs17 for both TIP and the \ion{Ca}{ii}\,IR camera. The spectral sampling of \ion{Ca}{ii}\,IR was 1.6\,pm per pixel, after a binning by two in wavelength to increase the light level. The step width was 0\farcs35 for all instruments in this case, and the slit width and spatial sampling for POLIS were the same as given above for observations Nos.~1 and 2. The integration time per scan step was 30\,seconds for TIP and 26\,seconds for POLIS, allowing us to observe also the weakest polarisation signals belonging to the weak magnetic fields in the internetwork of the QS because of the high signal-to-noise (S/N) ratio of the data. The rms noise of the polarisation signal at continuum wavelengths was about 3$\cdot 10^{-4}$\,of the continuum intensity $I_c$ at 630\,nm and 4.3$\cdot 10^{-4}$ of  $I_c$ for TIP at 1083\,nm. 

The PCO 4000 camera for the \ion{Ca}{ii}\,IR spectra was triggered by TIP at
the start of each scan step and used an exposure time of 25\,seconds, slightly
shorter than the integration time for TIP and POLIS. The spectra of
\ion{Ca}{ii}\,IR were taken through the polarimeter of TIP, including the
separation into two orthogonally polarised beams. In the data reduction, the
two beams in \ion{Ca}{ii}\,IR were only cut out from the image and added,
because without the synchronization  to the polarisation modulation of TIP
only intensity spectra can be obtained. This observation will be referred to
as No.~3 in the following. The wavelength range in \ion{Ca}{ii}\,IR extended from about 853.3\,nm up to 854.8\,nm thanks to the large CCD size \citep[compare to the range commonly used in IBIS, e.g.,][]{cauzzi+etal2008}.

With the same setup, a time series of about 50 minutes duration was obtained
on 8 September 2009 between UT 08:07:40 and 08:58:59 (observation No.~4). The integration time per scan step was 5 seconds and the observation consisted of 112 repeated scans of four steps of 0\farcs5 step width each. Only one of the four scan steps from the TIP and POLIS data was co-spatial in all wavelengths (396\,nm, 630\,nm, 854\,nm, 1083\,nm) because of the differential refraction in the Earths' atmosphere \citep{reardon2006,beck+etal2008} that was (partly) compensated by intentionally displacing the two slits of POLIS and TIP \citep[see][]{felipe+etal2010}. We therefore used only the co-spatial scan step that was observed with a cadence of about 27 seconds. 

Figure \ref{overview_2009} shows overview maps of the observations Nos.~3 and
4, derived from the spectra after the data reduction steps
described in the next section. The data have only been aligned with pixel
accuracy ($\sim 0\farcs36$), because a better alignment is not critical for
the present investigation and is also difficult to achieve for observation No.~3 with its
long integration time per scan step. For observation No.~3 ({\em left panel}),
most of the individual features can be identified in each wavelength range,
i.e., the strong polarisation signals of the photospheric \ion{Si}{i} line at
1082.7\,nm ({\em top left}) match those of the \ion{Fe}{i} lines at 630\,nm
({\em third column} in the {\em top row}) and also the emission in both
chromospheric \ion{Ca}{ii} lines matches ({\em second} and {\em fourth column}
in the {\em top row}). In the maps of the continuum intensity $I_c$ in the
{\em bottom row}, common features are more difficult to detect, but going from
one wavelength to the next allows one to re-encounter some of them. The most
prominent examples are four roundish darkenings, seen best at 630\,nm and
396\,nm ({\em white squares} in the $I_c$ maps). These darkenings re-appear in
the near-IR wavelengths as well, with reduced contrast and slightly
displaced. The corresponding maps of observation No.~4 are shown at the {\em
  right-hand side} of Fig.~\ref{overview_2009}. The maps of the continuum
intensity in the {\em lower row} match well to each other in this case, e.g.,
compare the locations of the roundish darkenings around $(t,y)=(45\,{\rm min},40^{\prime\prime})$.
\section{Additional data reduction steps\label{add_data}}
\subsection{Stray light correction for \ion{Ca}{ii}\,H}
In a recent study \citep[][BE11]{beck+etal2011a}, we
investigated the stray light contamination in POLIS. It turned out that for
the \ion{Ca}{ii}\,H channel two contributions are important: spectrally
undispersed stray light of about 5\,\% of amplitude created by scattering
inside POLIS and spectrally dispersed spatial stray light of about 20\,\%
amplitude. Following the procedure outlined in BE11, we corrected the
\ion{Ca}{ii}\,H spectra of POLIS for these two stray light contributions, using
a single stray light profile averaged over the full FOV. We subtracted first 5\,\% of the intensity in the line wing from all wavelengths and subsequently 20\,\% of the average profile from each individual profile in the observed FOV. 

\begin{figure}
\centerline{\resizebox{8.8cm}{!}{\includegraphics{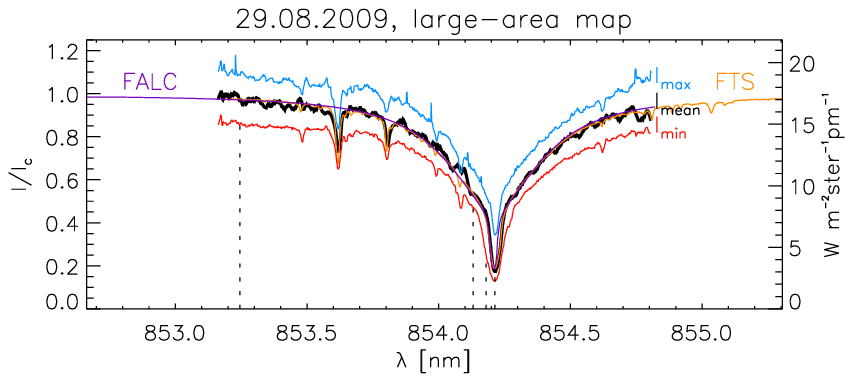}}\hspace*{.6cm}}
\centerline{\resizebox{8.8cm}{!}{\includegraphics{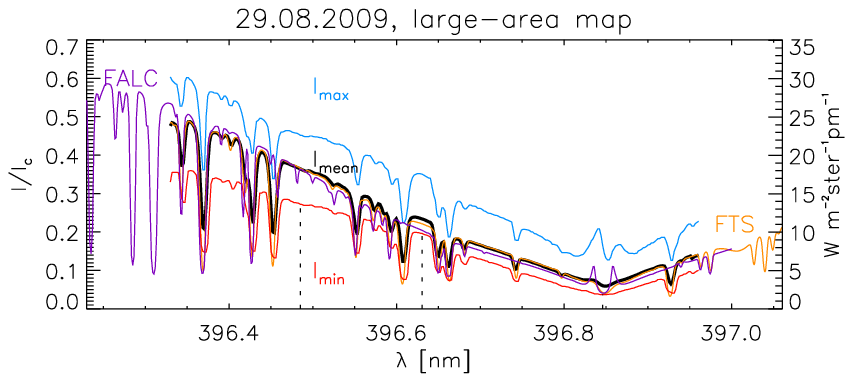}}\hspace*{.6cm}}
\caption{Average observed spectra  from the observation No.~3 in comparison to
  the FTS atlas ({\em orange}) and a synthetic FALC NLTE spectrum ({\em
    purple}). {\em Top}: \ion{Ca}{ii}\,IR at 854\,nm. {\em Bottom}: \ion{Ca}{ii}\,H at 396\,nm. {\em Black lines}: observed average profile. {\em Red/blue}: minimum/maximum observed intensity at any wavelength. The {\em vertical dotted lines} indicate the wavelengths used as examples later on.\label{int_norm}}
\end{figure} 
\subsection{Intensity normalization}
For 630\,nm, \ion{Ca}{ii}\,IR, and the TIP spectra, the intensity
normalization presents no problems because the observations were done on disc
centre and the spectral range covers continuum wavelength ranges. These
spectra therefore can be normalized to the continuum intensity, and can then be
directly compared to other theoretical or observed reference profiles. 

For the \ion{Ca}{ii}\,H spectra, no true continuum wavelengths exist inside the
observed spectral range. We thus used a (pseudo)continuum wavelength around
396.5\,nm for the normalization. We first determined the normalization
coefficient that brought the average observed spectrum on that of the FTS
atlas \citep{kurucz+etal1984,neckel1999}. To obtain an absolute intensity scale, we then synthesized \ion{Ca}{ii}\,H and \ion{Ca}{ii}\,IR spectra from the FALC atmosphere stratification \citep{fontenla+etal2006} with the NLTE code of \citet{uitenbroek2001}. The intensity of the NLTE profile is given in units of radiated energy per area, time, solid angle, and frequency interval and can be normalized to its ``continuum intensity'' using spectral windows without strong line blends far to the red and blue of the \ion{Ca}{ii}\,H line core. 

Figure \ref{int_norm} shows the final match of the average observed profiles
of \ion{Ca}{ii}\,H and the \ion{Ca}{ii}\,IR line at 854.2\,nm in observation No.~3, the
corresponding section of the FTS atlas, and the FALC NLTE spectrum in both relative intensities $I/I_c$ (scale {\em at left}) and absolute energy
units (scale {\em at right}). The corresponding profiles of observations
Nos.~1, 2, and 4 are displayed in Fig.~\ref{av_prof1} in Appendix
\ref{appa}. The shape of the lines is nearly identical in all average profiles
but for the very line core of \ion{Ca}{ii}\,H that shows some variation
between the different observations. This ensures that the conversion from the
observed intensity in detector counts to the absolute intensity provided by
the synthetic NLTE profile is reliable across different observations. In the following, we will usually denote the intensity as fraction of $I_c$; the conversion coefficient is 17.4\,Wm$^{-2}$ster$^{-1}$pm$^{-1}$ ($\equiv I_c = 1$) at 854\,nm \citep[cf.][their Fig.~1 yields about 17.3\,Wm$^{-2}$ster$^{-1}$pm$^{-1}$]{leenaarts+etal2009} and 50.1\,Wm$^{-2}$ster$^{-1}$pm$^{-1}$ at 396\,nm.

For each observation, we also overplotted the profiles corresponding to the
maximum and minimum intensity observed at each wavelength. These profiles
delimit the range of observed intensity variations, but do not correspond to
an individual observed profile. For the four observations of \ion{Ca}{ii}\,H,
the range of variation is very similar, about $\pm (0.1$\,--\,0.2) of $I_c$
around the average intensity; for the \ion{Ca}{ii}\,IR line at 854.2\,nm, the range is about $\pm 0.15$ of $I_c$.

Because all average profiles were found to be quite similar, we decided to
average all quantities and statistics over all available observations, i.e., all four for \ion{Ca}{ii}\,H and observations Nos.~3 and 4 for the \ion{Ca}{ii}\,IR line at 854.2\,nm. 
\subsection{Polarisation masks}
For all observations, simultaneous polarimetric data are available, either in
the two \ion{Fe}{i} lines at 630\,nm (all observations) or in the \ion{Si}{i}
line at 1082.7\,nm (observations Nos.~3 and 4). We defined masks delimiting locations with significant polarisation amplitude in the maps of the
wavelength-integrated unsigned Stokes $V$ signal \citep[for examples
see][their Appendix A]{beck+rammacher2010}. The threshold was lower than the one used in Fig.~\ref{overview_2009} where only the strongest polarisation signals are marked. With the masks, in each observed FOV three samples were defined: all locations with significant photospheric polarisation signal (``magnetic''), all locations without significant polarisation signal (``field-free''), and the full FOV. The magnetic sample covered about 10\,--\,20\,\% of the total FOV.
\section{Intensity statistics\label{sec_int}}
For the intensity statistics, we derived the distributions of intensities as function of the wavelength for the three samples inside the FOV defined above. We then determined the characteristic quantities of the intensity distributions up to second order (average value, standard deviation, skewness of distribution).
\subsection{Average profiles $I(\lambda)$}
Average profiles of \ion{Ca}{ii}\,H are discussed in detail in BE08. We thus
refer the reader to the latter publication, and only summarize their findings
here. The average profile of \ion{Ca}{ii}\,H for both field-free and magnetic
locations shows an asymmetric emission pattern near the line core, with
stronger emission in the H$_{2V}$ emission peak to the blue of the core at about 396.83\,nm than in H$_{2R}$ to the red at about 396.87\,nm (see Fig.~17 in BE08). The asymmetry of the two emission peaks is the signature of upwards propagating shock waves. In the most quiet parts of the FOV, the average profile shows a nearly reversal-free shape without emission as discussed in \citet{rezaei+etal2008}. The average profile of magnetic locations differs from that of the field-free locations by an additional contribution to the emission that is roughly symmetric around the very line core. This contribution presumably indicates a chromospheric heating process that has no counterpart in a Doppler shift of the line in observations on disc centre (e.g, magnetic reconnection with horizontal outflows or transversal wave modes). 
\begin{figure}
\centerline{\resizebox{8.8cm}{!}{\includegraphics{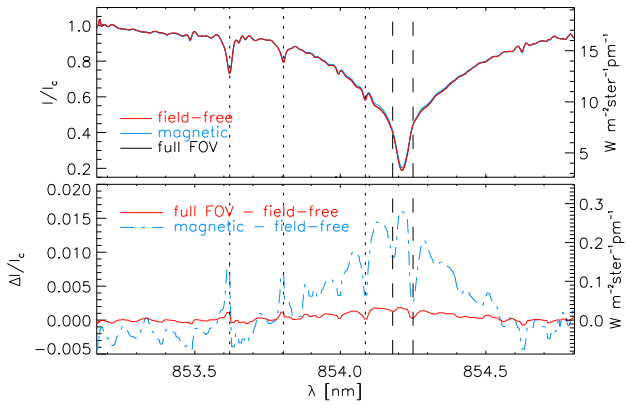}\hspace*{.6cm}}}
\caption{Average profiles of \ion{Ca}{ii}\,IR at 854\,nm. {\em Top}: average profiles of the full FOV ({\em black}), field-free ({\em red}) and magnetic locations ({\em blue}). {\em Bottom}: difference of the profiles of the full FOV and the field-free locations ({\em red solid}) and difference of magnetic and field-free locations ({\em blue dash-dotted}). The {\em  dotted vertical lines} denote the locations of line blends, the {\em dashed vertical lines} an asymmetry in the intensity differences.\label{av_prof_caiir}}
\end{figure}

Whereas the shape of the \ion{Ca}{ii}\,H line core changes by a few percent of
$I_c$ between the different samples (full FOV, magnetic and field-free
locations), the variability of the \ion{Ca}{ii}\,IR line at 854.2\,nm is smaller (Fig.~\ref{av_prof_caiir}). For the average profiles, one cannot really
distinguish between the magnetic locations, the field-free locations, or the
full FOV ({\em upper panel} of Fig.~\ref{av_prof_caiir}). Subtracting the
average profile of the field-free locations as the first-order estimate of the
most quiet regions with the lowest intensity reveals that the average profile
from magnetic regions exceeds it by about 1.5\,\% of $I_c$ near the line core
({\em lower panel} of Fig.~\ref{av_prof_caiir}). The difference is symmetrical
around the very line core on a large scale. The (shock) wave signature
presumably is reflected by the asymmetry of the difference intensity just to
the blue and the red of the line core ({\em dashed vertical lines} in the {\em
  lower panel} of Fig.~\ref{av_prof_caiir}), with a larger intensity to the
blue than to the red of the line core. The intensity differences at the cores
of line blends ({\em dotted vertical lines}) flip the orientation relative to
the neighboring wavelengths: the cores of the lines in the far line wing (e.g., at 853.62\,nm and 853.805\,nm) have an excess intensity over the neighboring wavelengths, whereas the blends near the core (e.g., at 854.086\,nm) have a lower intensity. This could be caused by either the temperature stratification or the different magnetic sensitivity of the respective transitions.
\subsection{Intensity distributions of $I(\lambda)$}
Figure \ref{individ_hists} displays the histograms of the intensity at a few selected wavelengths (marked by {\em vertical dotted lines} in Fig.~\ref{int_norm}, see Rezaei et al.~2007a or BE08 for more examples). The intensity distributions at each wavelength point for the three samples inside the FOVs do not differ much, but some tendencies can be discerned. The distributions for the magnetic locations are slightly broader than the other two, with a tail of the distribution towards high intensities. In the \ion{Ca}{ii}\,H line, the distribution around the H$_{2V}$ emission peak is significantly broader than that of the H$_{2R}$ emission peak. The distributions show a variation in their width that does not change monotonically with wavelength. The next two sections will quantify these findings. Table \ref{tab_stat} in Appendix \ref{appc} lists all parameters of the intensity distributions at a few selected wavelengths.
\subsection{Standard deviation of $I(\lambda)$}
The standard deviation $\sigma$ (or the rms variation, respectively) of the intensity distributions are shown in Fig.~\ref{fig_rms}. The {\em bottom panels} show the absolute rms values, i.e, the rms normalized to the continuum intensity $I_c$, as function of wavelength, the {\em middle panels} the relative rms, i.e., the rms normalized to the wavelength-dependent intensity of the spatially averaged profile $<I(\lambda)>$. 
\begin{figure}
\centerline{\resizebox{8.8cm}{!}{\includegraphics{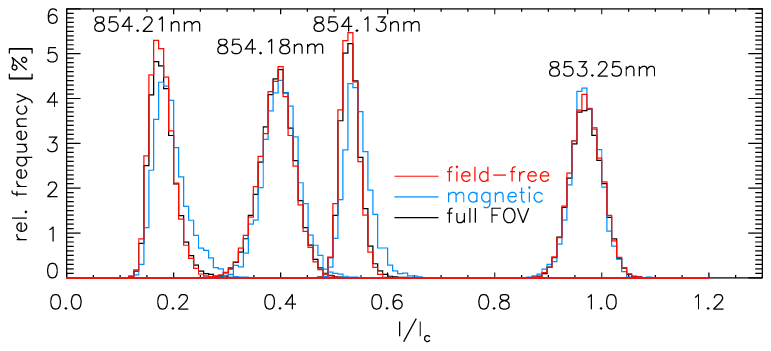}}}\vspace*{-.5cm}
\centerline{\resizebox{8.8cm}{!}{\includegraphics{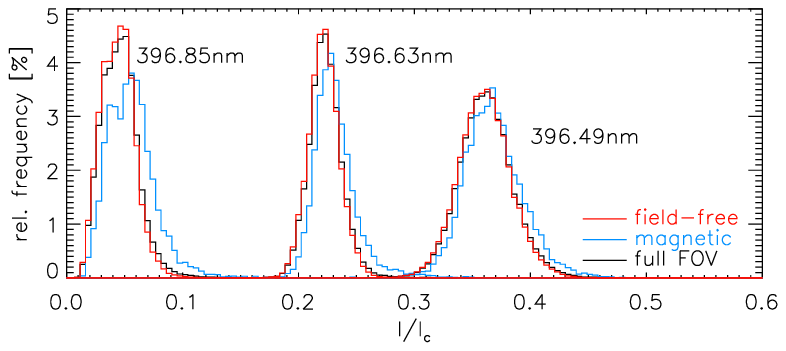}}}
\caption{Intensity histograms at selected wavelengths (denoted next to each histogram). {\em Top}: \ion{Ca}{ii}\,IR at 854\,nm. {\em Bottom}: \ion{Ca}{ii}\,H. {\em Black/blue/red lines}: full FOV, magnetic locations, and field-free locations.\label{individ_hists}}
\end{figure}
\begin{figure*}
\centerline{\hspace*{-1.cm}\resizebox{8.8cm}{!}{\includegraphics{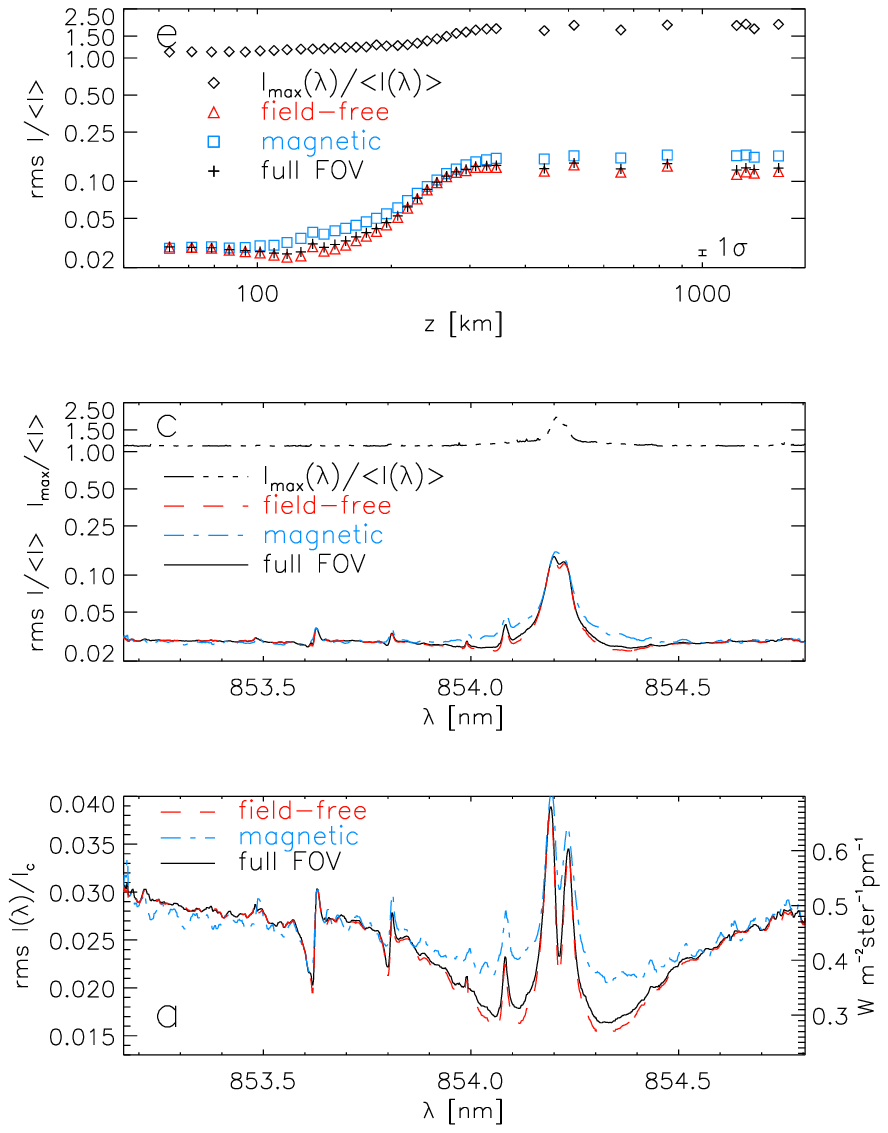}}\hspace*{.4cm}\resizebox{8.8cm}{!}{\includegraphics{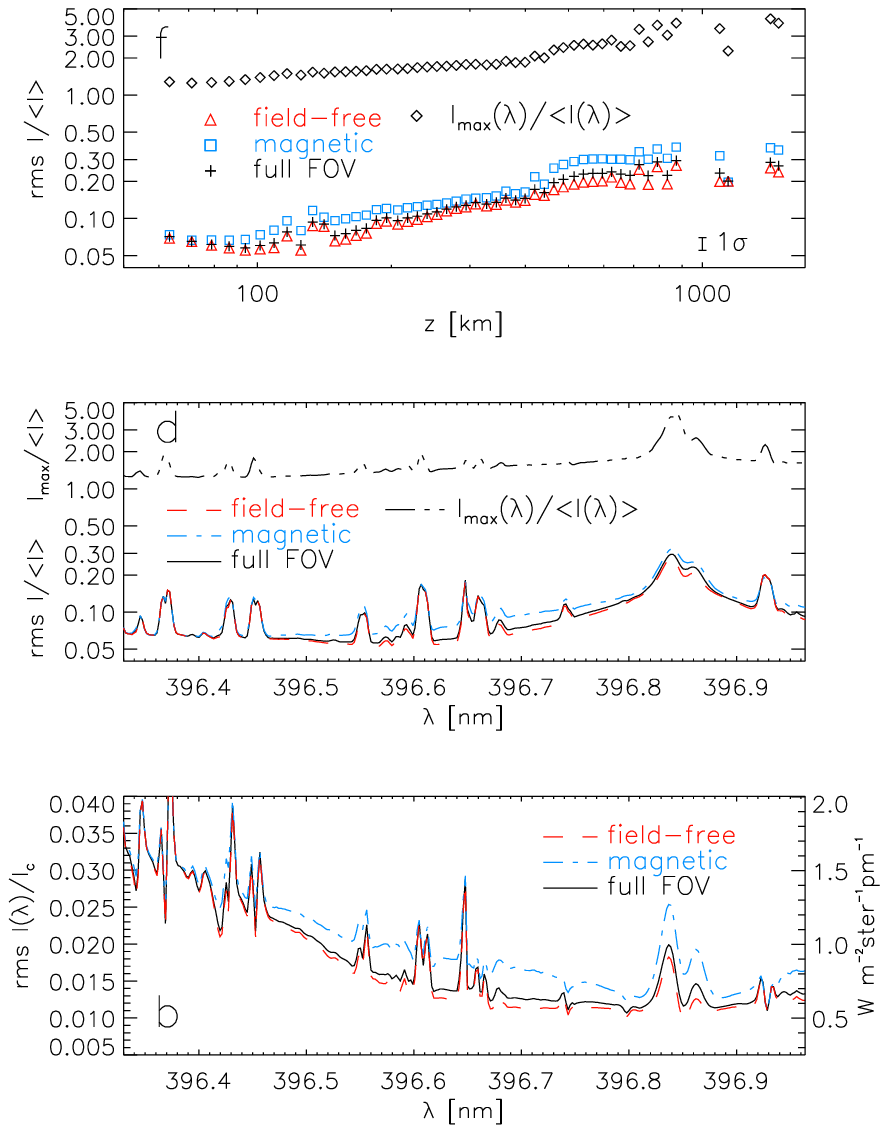}}}
\caption{Absolute ({\em bottom row}) and relative ({\em middle row}) rms variation of the intensity as function of wavelength for \ion{Ca}{ii}\,IR at 854\,nm ({\em left}) and  \ion{Ca}{ii}\,H ({\em right}). The {\em solid black lines} denote the full FOV, the {\em red dashed} and {\em blue dash-dotted lines} the field-free and magnetic locations, respectively. {\em Top row}: relative fluctuations {\em vs.}~geometrical height. {\em Black crosses}: full FOV. {\em Blue squares}: magnetic locations. {\em Red triangles}: field-free locations. The ratio of maximum to average intensity is overplotted in the {\em upper two rows} with a {\em black dashed triple-dotted line} and {\em black diamonds}, respectively. The short bars in the {\em lower right corners} in panels {\em e} and {\em f} indicate the average range of scatter for the data points plotted. \label{fig_rms}}
\end{figure*}

The absolute rms values  for both lines (panels {\em a} and {\em b} of
Fig.~\ref{fig_rms}) share some common trends. The rms variation reduces from the low-forming (pseudo)continuum wavelength ranges in the far line wing towards the line core, with profound minima for \ion{Ca}{ii}\,IR around 854.1\,nm and 854.35\,nm, respectively. Near the line cores, the absolute rms values increase again strongly. For the \ion{Ca}{ii}\,IR line at 854.2\,nm, the maximum absolute rms value is reached near the line core, for \ion{Ca}{ii}\,H in the line wing. This should be related to the significantly steeper line profile of the \ion{Ca}{ii}\,IR line, where Doppler shifts of the absorption profile will yield large variations of the intensity values taken at a fixed wavelength. The large rms fluctuations in the line wing of \ion{Ca}{ii}\,H are caused by the contrast of the granulation pattern, coupled with the high sensitivity to temperature at the near-UV wavelength. For both lines, the field-free locations show the smallest rms variations at almost all wavelengths. On magnetic locations, the rms value reduces less than for the other two samples in an intermediate range from 853.8\,nm (396.45\,nm) to the line core. Near the very line core, the \ion{Ca}{ii}\,H line has a clear asymmetry between the wavelengths of the red and blue emission peaks H$_{2R}$ and H$_{2V}$, with larger rms variations for H$_{2V}$. The rms values of \ion{Ca}{ii}\,IR just to the red and blue of the line core mimick this asymmetry, even if the average intensity profile of the latter does not show prominent emission peaks.

The relative rms values (panels {\em c} and {\em d} of Fig.~\ref{fig_rms})
behave quite similar to the absolute rms values, but the minimum of rms
fluctuations is much less pronounced for the \ion{Ca}{ii}\,IR line at 854.2\,nm
and nearly completely vanishes for the magnetic locations ({\em blue line} in
{\em left middle panel}). The relative rms variations of the intensity near
the line core are about 20\,\% (30\,\%) for \ion{Ca}{ii}\,IR (\ion{Ca}{ii}\,H). The ratio of maximal and average intensity ({\em dash-dotted line} in panels {\em c} and {\em d} of Fig.~\ref{fig_rms}) reaches up to a factor of two for \ion{Ca}{ii}\,IR and about four for \ion{Ca}{ii}\,H, i.e., the intensity can more than double relative to the average value. Because the intensity histograms start to be asymmetric for wavelengths close to the line core (Fig.~\ref{individ_hists}), the rms value presumably slightly underestimates the full amount of the fluctuations, which also do not have the same range towards higher or lower than average intensities.

The {\em upper panels} {\em e} and  {\em f} of Fig.~\ref{fig_rms} show the
relative rms fluctuations {\em vs.}~geometrical height. The conversion from wavelength to geometrical height was done by attributing the optical depth value at the center of the intensity response function to each wavelength, and then using the geometrical height corresponding to that optical depth in a reference atmosphere model \citep[see Appendix \ref{appb}, or also][]{leenaarts+etal2010}. Each observed spectral range covered several hundred wavelength points, but especially in the line wing several wavelengths are attributed to the same or a very similar geometrical height (cf.~the {\em lower panel} of Fig.~\ref{cair_response}). To avoid overcrowding, we plotted only the average value of the rms fluctuations on all unique height points (about 70 values). As estimate of the scatter at one given height, we averaged the range of scatter on all of the data points plotted (short bars in the {\em lower right corners} of panels {\em e} and  {\em f}). The scatter at some specific heights can be much larger (cf.~the {\em right panel} of Fig.~\ref{fig_rmst_comp} later on). In the plots of the relative rms fluctuations {\em vs.}~geometrical height ({\em top panels} of Fig.~\ref{fig_rms}), the fluctuations monotonically increase with height after passing through a minimum between 100\,km and 200\,km height. They level off to a plateau at about 500\,--\,700\,km height.
\begin{figure*}
\centerline{\hspace*{-.8cm}\resizebox{8.8cm}{!}{\includegraphics{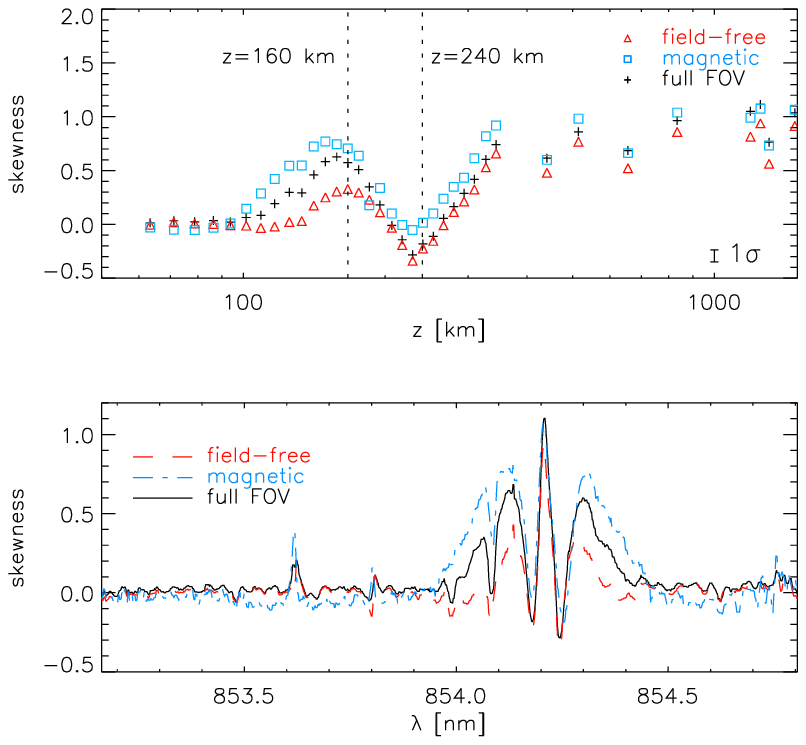}}\hspace*{-.0cm}\resizebox{8.8cm}{!}{\includegraphics{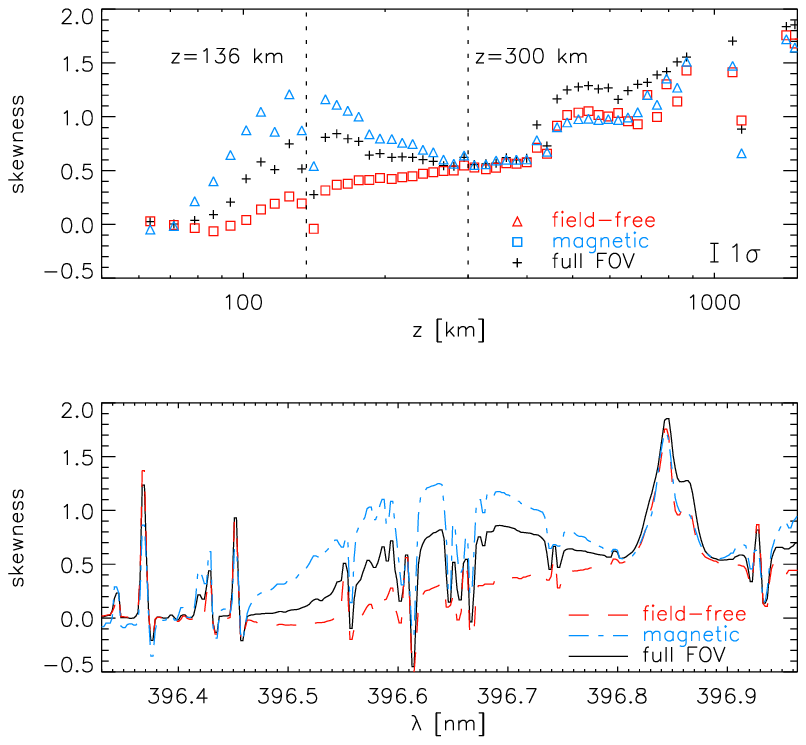}}}
\caption{Skewness as function of wavelength ({\em bottom}) and {\em vs.}~geometrical height ({\em top}) for \ion{Ca}{ii}\,IR at 854\,nm ({\em left}) and  \ion{Ca}{ii}\,H ({\em right}). Same labels and notation as in Fig.~\ref{fig_rms}.\label{fig_skew}}
\end{figure*}
\begin{figure*}
\resizebox{17.6cm}{!}{\includegraphics{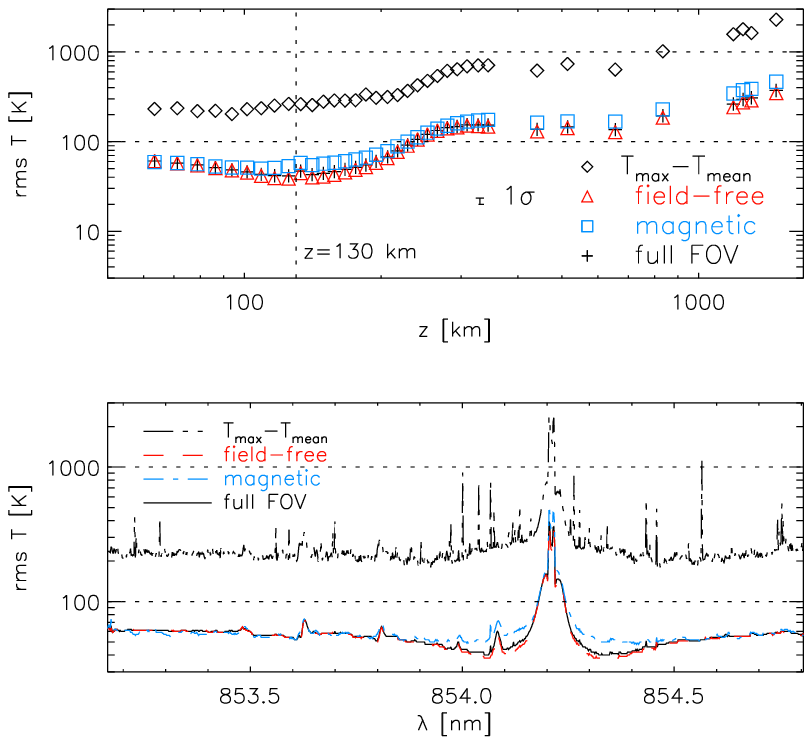}\includegraphics{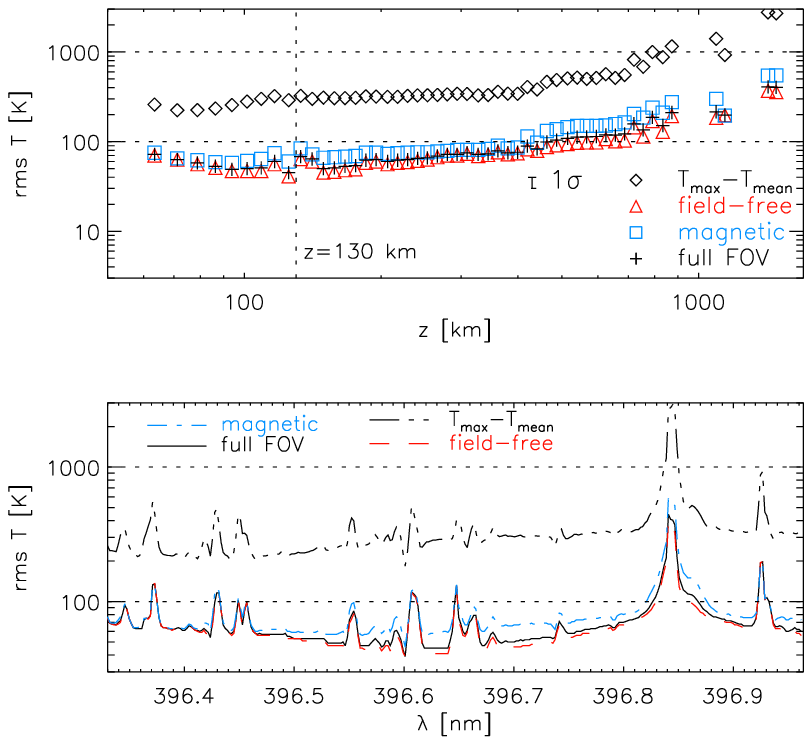}}
\caption{Rms fluctuations of brightness temperature {\em vs.}~wavelength ({\em bottom row}) and geometrical height ({\em top row}). {\em Left column}: \ion{Ca}{ii}\,IR at 854\,nm. {\em Right column}: \ion{Ca}{ii}\,H at 396\,nm. The difference of maximum and average temperature is overplotted with a {\em black dashed triple-dotted line} and {\em black diamonds}, respectively. Other labels and notation are as in Fig.~\ref{fig_rms}.\label{fig_rmst}}
\end{figure*}
\subsection{Skewness of distributions of $I(\lambda)$}
The skewness of the intensity distributions is defined as 
\begin{eqnarray}
{\rm Skewness} = \frac{1}{N} \sum_{j=1}^{N} \left( \frac{x_j-\bar{x}}{\sigma}\right)^3  \;,
\end{eqnarray}
where $N$ is the total number of sample points $x_i$ and $\sigma$ the standard deviation \citep[][Chapter 14]{press+etal1992}. 

The individual histograms in Fig.~\ref{individ_hists} already hinted at a
variation of the skewness with wavelength as well as between the three
samples. Figure \ref{fig_skew} shows the skewness as function of
wavelength ({\em lower panels}) and {\em vs.}~geometrical height ({\em upper
  panels}). The differences between the three samples are similar to that in
the rms variations, with the magnetic field locations showing largest and the
field-free locations the smallest values, but they are much more pronounced here. In a wavelength range intermediate between line wing and core
(853.9\,--\,854.15\,nm and 396.45\,--396.75\,nm), the magnetic locations
exhibit a skewness larger by about a factor of two than the field-free
locations, with the skewness of the full FOV in the middle between the
others. This large difference is maintained up to close to the line core,
where the skewness becomes nearly identical again. 

The maximal difference of the skewness between magnetic and field-free locations is reached at wavelengths that form at geometrical heights of about 150\,km for both spectral lines ({\em top panels} of Fig.~\ref{fig_skew}). At a height of about 250\,--\,300\,km, the skewness values equalize again. Both spectral lines contain several line blends that could be the reason for the difference in skewness between the three samples in the FOV, but the same plot as in the {\em upper panels} of Fig.~\ref{fig_skew} using only wavelengths outside of line blends yielded the same behaviour. Doppler shifts of the whole line profile presumably can also be excluded as a reason for the difference in skewness, because for instance for \ion{Ca}{ii}\,H the slope of the line profile is rather small in the wavelength region of the largest difference in skewness. It thus indicates a
significant difference of the intensity distributions between the outer wing and the line core on locations with and without magnetic fields.  
\begin{figure*}
\centerline{\resizebox{17.6cm}{!}{\includegraphics{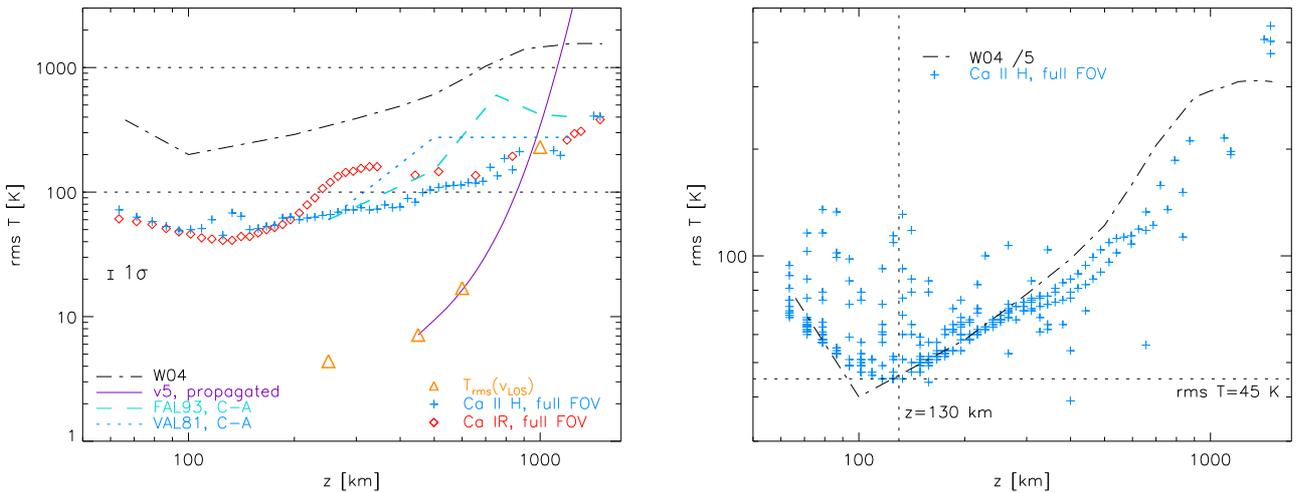}}}
\caption{Rms fluctuations of temperature {\em vs.}~geometrical height in comparison to other measurements ({\em left}). {\em Orange triangles}: temperature fluctuations corresponding to observed rms LOS velocity (BE09). {\em Purple solid line}: propagated LOS velocity of \ion{Fe}{i} at 396.6\,nm converted to the corresponding temperature equivalent (BE09). {\em Black dash-dotted}: rms of the kinetic temperature in the 3D HD simulation of WE04 as read off from their Fig.~9. {\em Blue dotted line}: temperature difference of models C and A in VAL81. {\em Turquoise dashed line}: the same for FAL93. {\em Red diamonds/blue crosses}: temperature fluctuations corresponding to the intensity of \ion{Ca}{ii}\,IR and \ion{Ca}{ii}\,H, respectively. {\em Right}: temperature fluctuations corresponding to the intensity of \ion{Ca}{ii}\,H ({\em blue crosses}) and rms of the kinetic temperature of WE04 divided by 5 ({\em dash-dotted line}). \label{fig_rmst_comp}}
\end{figure*}
\subsection{Conversion from intensity to temperature fluctuations}
The observed (absolute) intensities cannot directly be converted to corresponding gas temperatures without solving the radiative transfer equation, but the relative intensity fluctuations can be approximately converted to {\em brightness} temperature fluctuations using the Planck curve. The intensity response functions (Appendix \ref{appb}) provide a mapping from a wavelength $\lambda$ to a corresponding optical depth $\tau(\lambda)$ -- taken to be the center of the intensity response function -- and hence, to the average temperature  $T(\tau(\lambda))$ as given in a reference atmosphere model. If one assumes that the average intensity at one wavelength $I_{\rm mean}(\lambda)$ corresponds to the temperature $T_0(\tau)$ predicted by the reference atmosphere model for the respective optical depth $\tau(\lambda)$, the deviation $\Delta I$ from the average intensity value can be attributed to a variation $\Delta T$ of the temperature. The modulus of the temperature variation can be estimated using the Planck curve (Appendix \ref{appd}). We used the Harvard Smithsonian Reference Atmosphere model \citep[HSRA,][]{gingerich+etal1971} as temperature reference.

Figure \ref{fig_rmst} shows the resulting rms fluctuations of brightness
temperature as function of wavelength and geometrical height for the
\ion{Ca}{ii}\,IR ({\em left column}) and the \ion{Ca}{ii}\,H line ({\em right
  column}), respectively. The rms fluctuations of brightness temperature show a minimum of about 45\,K at heights of about 130\,km and do not exceed 100\,K up to a height of about 300\,km. They reach values of about 500\,K at 1\,Mm height. The maximal deviation T$_{\rm max}$-T$_{\rm mean}$ ({\em dotted lines} and {\em black diamonds} in Fig.~\ref{fig_rmst}) from the reference model is about 200\,K in the line wing (or lower photosphere), increases nearly monotonically with height in the atmosphere, and exceeds 1000\,K at $z\sim 800\,$km. Because of the increasing deviations from LTE with height in the solar atmosphere, the conversion should differ successively more from the kinetic temperature the higher a given wavelength forms in the solar atmosphere. 

Figure \ref{fig_rmst_comp} compares the rms fluctuations of brightness
temperature derived from the intensity rms ({\em red diamonds} for \ion{Ca}{ii} IR and {\em blue crosses} for \ion{Ca}{ii} H) with other results for the {\em gas} temperature rms in the solar atmosphere. As two examples of the variation in semi-empirical static temperature models, we chose the temperature difference between 
the models C (average internetwork area) and A (chromospheric faint location
in quiet Sun) of \citet[][VAL81, {\em blue dotted line} in
Fig.~\ref{fig_rmst_comp}]{vernazza+etal1981} and \citet[][FAL93, {\em turquoise dashed line} in Fig.~\ref{fig_rmst_comp}]{fontenla+etal1993}, following the
approach of \citet{kalkofen2012}. From the first paper of this series (BE09),
we took the rms temperature fluctuations corresponding to observed rms line-of-sight (LOS) line-core velocities. The energy content of the rms LOS velocities was converted to the corresponding enhancement of gas temperature by equating the increase of the internal energy and the mechanical energy ({\em orange triangles}). In addition to the observed rms velocities, we also
overplotted the curve obtained by propagating the LOS velocity of the
\ion{Fe}{i} line at 396.6\,nm to higher layers in the limit of linear
perturbation theory and subsequently converting the resulting rms velocities
again to the equivalent temperature variation ({\em purple solid line}). Finally, the {\em black dash-dotted line} shows the rms of the {\em kinetic} temperature in a numerical 3D HD simulation done with the CO$^5$BOLD code \citep{freytag+etal2012} as given in Fig.~9 of \citet[][WE04]{wedemeyer+etal2004}. The latter matches the shape of the temperature rms curve derived from the \ion{Ca}{ii}\,H intensity surprisingly well, but is a factor of about five larger ({\em right panel}). In this case, the temperature rms fluctuations of all 326 wavelengths points of the \ion{Ca}{ii} H spectra were plotted individually without averaging. The increased scatter below a height of about 200\,km is caused by wavelengths in the wing and core of the line blends.

The rms variations from all different approaches share, however, the following
properties: a) a local minimum of rms fluctuations at about $z=130$\,km, b) a steepening of the curves at about $z=500$\,km, and c) rms
fluctuations of about 300\,--\,400\,K at about $z=1000$\,km. The energy
equivalent of the rms LOS velocities below $z=1000$\,km falls short by an
order of magnitude or more in comparison to the other values. The difference
between the ``cool'' and average semi-empirical models (VAL/FAL C$-$A) matches
the rms variation found from the intensity fluctuations around $z=1000$\,km. 

\section{Summary and Discussion\label{summ_disc}}
\subsection{Summary}
The spectral lines of \ion{Ca}{ii}\,H and \ion{Ca}{ii}\,IR at 854.2\,nm cover
smoothly formation heights from the continuum forming layers of the solar
atmosphere in the line wing to the lower chromosphere in the line core. This
offers the possibility to obtain information on the solar atmosphere not only
from absolute values of observed quantities, but also already from their
relative variation throughout the lines. As far as the intensity statistics
are concerned, the two lines behave nearly identical, with the following
common characteristics: 1.~a general increase of the relative rms fluctuations
$\sigma_{I}/I_{\rm mean}$ from the line wing to the line core, with minimal rms at wavelengths forming in the low photosphere at about 130\,km height. The range of fluctuations is between 3\,\% to 30\,\% rms and up to 400\,\% maximum variation at the line core. 2.~Larger rms fluctuations on locations with magnetic fields than on field-free locations. 3.~A variable skewness of the intensity distributions caused by extended tails towards high intensities. 4.~A pronounced local maximum of the skewness for locations with magnetic fields at intermediate wavelengths of 853.9\,--\,854.15\,nm and 396.45\,--396.75\,nm. 5.~The
temperature difference  between the original and a modified HSRA model
without chromospheric temperature rise ($\equiv$ an atmosphere in radiative
equilibrium) is about 1000\,K at a height of 1\,Mm (Fig.\,13, BE09). The rms variations of temperature from either the velocity or the intensity fluctuations do not reproduce such an average temperature enhancement because they suffice only for an enhancement of at maximum about 500\,K. We remark that the observations had a spatial resolution of about 1$^{\prime\prime}$.
\subsection{RMS of intensity and temperature fluctuations}\label{sec:disc_rms}
After passing through a minimum of the fluctuations at about 130\,km height ($\log \tau \approx -1$) caused by the reduction of the granulation contrast \citep[cf.~][]{puschmann+etal2003,puschmann+etal2005}, the rms fluctuations indicate an increase with height in the solar atmosphere that is expected for upwards propagating acoustic waves with a steepening of the wave amplitude caused by the exponentially decreasing gas pressure. The fluctuations are larger on locations co-spatial to photospheric magnetic fields. To evaluate the significance of the differences in the mean and rms variations between the magnetic and field-free samples (Figs.~\ref{individ_hists} and \ref{fig_rms}, and Table~B.1), we used a \emph{t-Student test} and \emph{F-test} \citep{press+etal1992}. These tests indicated that the differences are significant by a large margin.  The rms fluctuations amount to 61\,W/m$^{-2}$ster$^{-1}$ in a 0.1\,nm
window around the \ion{Ca}{ii}\,H line core and to 31\,W/m$^{-2}$ster$^{-1}$
for a similar wavelength window around the \ion{Ca}{ii}\,IR line core, where
the maximal variations can be much larger than that. 

We converted the relative intensity fluctuations to the corresponding
brightness temperature variation using the Planck curve, implicitly assuming
LTE in the approach. This yielded rms fluctuations of the brightness
temperature below 100\,K for the photosphere up to $z\sim 300$\,km, and up to
500\,K in the lower chromosphere at $z\sim 1$\,Mm. The trend of brightness
temperature fluctuations with height derived from the intensity at different
wavelengths in the \ion{Ca}{ii}\,H line matches that in the rms fluctuations
of the kinetic temperature in the 3D HD simulations of WE04, but the latter values are larger by a factor of about five. Part of this discrepancy can be
attributed to the spatial resolution of about 1$^{\prime\prime}$ of the data
used here. Improving the spatial resolution from 1$^{\prime\prime}$ to,
for instance, the 0\farcs32 of the Hinode/SP \citep{kosugi+etal2007} increases
the rms variations by a factor of about 2\,--\,3 \citep[][Table
3]{puschmann+beck2011}. For \ion{Ca}{ii} IR, data at higher spatial resolution than ours can currently be obtained with IBIS at the Dunn Solar Telescope or CRISP at the Swedish Solar Telescope \citep{scharmer+etal2008}. The GREGOR Fabry-Per{\'o}t Interferometer \citep[cf.][and references therein]{puschmann+etal2011} will provide another source of \ion{Ca}{ii} IR spectra at 0\farcs14 in the near future. The match of our observed temperature rms and the recently published temperature rms in numerical simulations in \citet[][{\em upper right panel} of their Fig.~9]{freytag+etal2012} would presumably be better than the match with WE04 already without any artificial down-scaling of the fluctuations in the simulations, but their simulations cover only a smaller height range compared to WE04.
\subsection{Skewness of intensity distributions}\label{sec:disc_skw}
A (large) positive skewness of a probability distribution indicates an
extended tail of values above the mean value. The skewness as characteristic quantity, however, has to be taken with care because even a perfectly
symmetric Gaussian distribution can yield a non-zero skewness in a limited
sample. The standard deviation of the skewness for a normal distribution is
given approximately by $\sqrt{15/N}$, where $N$ denotes the number of sample
points \citep{press+etal1992}. In our case, $N\,\approx 10.000$ for the
smallest of the samples in the FOV, which yields a standard variation of the
skewness of about 0.04. Hence, all measured values of skewness
(Fig.~\ref{fig_skew}) above about 0.1 should be significant. 

The skewness of the observed intensity distributions is generally positive, close to zero in the line wing/lower heights and up to two near the line core. This implies a more frequent occurrence of excursions towards high intensities, which is expected for acoustic waves forming ``hot'' shock fronts, not ``cold'' ones. The positive skewness would also fit to the existence of a limiting energetically lowest state of the atmosphere, e.g., given by an atmosphere in radiative equilibrium or with a stationary chromospheric temperature rise \citep[\emph{``basal energy flux''},][]{schrijver_95,schroeder+etal2012}, that intermittently is raised to a transient state of higher energy content. The basal flux would prevent the temperature, and hence the intensity from decreasing below some certain critical level, similar to the effect of a thermostat set to prevent freezing. On the upper end of the intensity range, the basal flux would, however, not impose any limitations on the maximum possible intensity value \citep[cf.~also][]{rezaei+etal2007}. In the presence of a basal energy flux one therefore would not expect to find a negative skewness for chromospheric intensities in the QS because the intensity distribution should have a (rigid) lower limit that cannot be exceeded, but no upper limit (cf., e.g., the intensity distribution at 854.21\,nm in the {\em upper panel} of Fig.~\ref{individ_hists}). It would be interesting to evaluate the skewness of emergent intensities in numerical simulations for comparison to our observational findings. 

The source of the observed skewness values in the intermediate wavelength
  range (Fig.~\ref{fig_skew}) remains elusive. High-intensity events at
  heights below 300\,km cause a distinct skewness pattern on magnetic regions
  compared to the field-free area. These events should be different from the
  shock fronts of steepening acoustic waves found at about 1\,Mm that
  induce the asymmetry of the intensity distributions near the line cores. At
  a height of 300\,km acoustic waves are still in the linear regime without
  strong damping by, e.g., radiative losses. Section \ref{sec:radlos} below
  discusses some possible processes that could modify the skewness in the
  presence of magnetic fields. The remarkably large skewness at the very line core of both lines, i.e., $z\,\approx\,1$\,Mm, originates from intensity brightenings caused by hot shock fronts at those heights. Unlike in the line wing, there is no difference of skewness between the magnetic and field-free locations in the line core. Hence, our results indicate rather different origins for the enhanced skewness in the wing and the core of the two lines.
\subsection{Assumptions and uncertainties\label{err_sec}}
\begin{figure}
\includegraphics{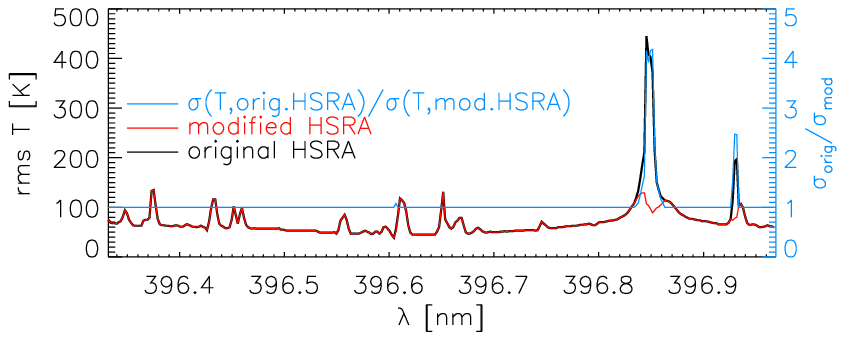}\vspace*{-.5cm}
\includegraphics{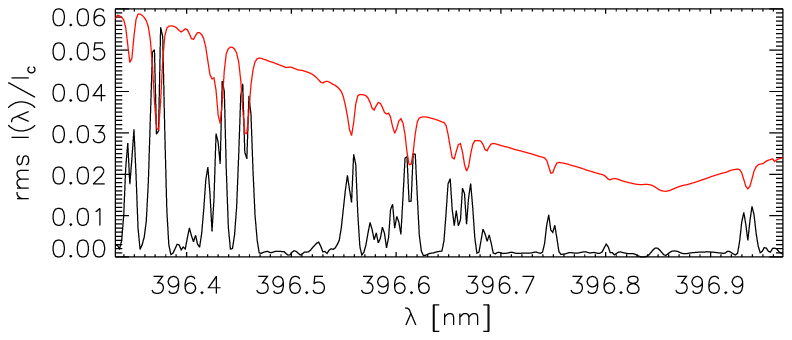}\vspace*{-.5cm}
\includegraphics{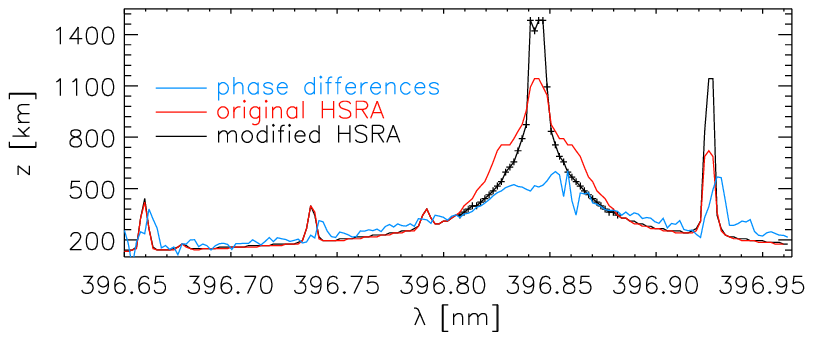}
\caption{Derivation of error estimates for \ion{Ca}{ii}\,H. {\em Bottom}: formation heights in the original ({\em red}) and modified HSRA model ({\em black}). The {\em blue line} gives the formation height from phase differences. {\em Middle}: rms fluctuations induced by random Doppler shifts ({\em black}). The {\em red line} shows an average profile. {\em Top}: temperature rms using the original ({\em black}) and modified HSRA model ({\em red}). Their ratio is given by the {\em blue line} with the axis on the right-hand side. \label{err_est}}
\end{figure}

The discrepancy between the observed and model temperature variations (Fig.~\ref{fig_rmst_comp}) depends on the rather long list of (strong) assumptions used in the derivation of the temperature fluctuations from the observed spectra. We try to estimate the error from different steps in the conversion in the following.
\paragraph{Relation between wavelengths and formation heights:}
We attributed a specific geometrical height to a given wavelength using the
intensity response functions in a static model atmosphere, ignoring both the
variable width of the formation height at each wavelength and the dynamical
character of the solar chromosphere. To test at least the dependence of the
formation height on the selected model atmosphere, we repeated the calculation
for the original and the modified HSRA model for the \ion{Ca}{ii}\,H line ({\em
  bottom panel} of Fig.~\ref{err_est}). Because the two model atmospheres are
identical up to $\log \tau \sim -4$, only wavelengths near the line core
change ({\em black crosses} in the {\em bottom panel} of Fig.~\ref{err_est}). Using the original HSRA model slightly reduces the formation height of the line core and raises a bit that of the two emission peaks. The modified HSRA model matches, however, the formation heights derived from the slope of phase differences (see BE08) around $\lambda \sim 396.82$\,nm and $396.87$\,nm better than the original HRSA model. The measurements of the phase differences are done in a dynamical atmosphere and therefore take spatial and temporal variations of the photospheric and chromospheric structure into account. The uncertainty in the formation height would slightly compress or expand all plots {\em vs.}~geometrical height, but would not change the general slopes of the curves. The derivation of formation heights from phase differences could be repeated for the \ion{Ca}{ii}\,IR line at 854.2\,nm with the data set No.~4.
\paragraph{Effect of Doppler shifts:}
The intensity fluctuations attributed to a specific wavelength could also have
some intrinsic flaw. Part of the scatter in intensity could come from
large-scale macroscopic Doppler shifts that displace the complete absorption
profile. For the data used here, this should, however, mainly affect the
\ion{Ca}{ii}\,IR line because of its steeper absorption profile near the line
core. For the \ion{Ca}{ii}\,H spectra from POLIS with a spectral sampling of
1.92\,pm -- corresponding to a velocity dispersion of about 1.5\,kms$^{-1}$
per pixel -- and given the mild slope in the line wing of the \ion{Ca}{ii}\,H
line, the intensity scatter outside any line core should only reflect true changes of the intensity at a given wavelength and not intensity variations caused by Doppler displacements of the line. 

As a test, we created a statistical sample of 50.000 \ion{Ca}{ii}\,H profiles by Doppler shifting an average observed profile with random Gaussian velocities with an rms value of 1\,kms$^{-1}$. The resulting intensity rms ({\em middle panel} of Fig.~\ref{err_est}) is close to zero outside of all line blends, but around each line blend the Doppler shifts induce a significant rms intensity fluctuation comparable to the observed values ({\em bottom right panel} in Fig.~\ref{fig_rms}). These additional fluctuations caused by Doppler shifts presumably are reflected in, e.g., the increased scatter between 50 and 150\,km height in the {\em right panel} of Fig.~\ref{fig_rmst_comp}, whereas the wavelengths outside the line blends yield the majority of the points that follow a rather smooth curve of lower rms fluctuations. We note that Doppler shifts cannot be the cause of the positive skewness of the intensity distributions, but broaden the intensity histograms because of the similar occurrence rates of blue and red shifts.
\paragraph{Reference model atmosphere and LTE assumption:} The most critical
effect in the derivation of the temperature fluctuations is of course the LTE
assumption, but also the choice of the reference model and the lack of treatment of the complete line formation in the radiative transfer have an effect on the retrieved temperature fluctuations. The latter is partly compensated because only relative variations are considered, i.e., the variation of the line shape (temperature gradient) compared to the average shape. For the conversion from intensity to temperature fluctuations, we tested the effect of selecting a specific temperature stratification by using the original or the modified HSRA model ({\em top panel} of Fig.~\ref{err_est}). The solar atmosphere presumably fluctuates roughly between these two cases, with the modified HSRA as lower boundary case. Throughout most of the line wing, the rms stays again unchanged because only the upper layers of the HSRA model were modified. The temperature rms fluctuations in the line core reduce by a factor of up to four when the modified version of the HSRA is used. The reason is that to first order the relative intensity variation is proportional to the relative temperature change times $T^{-1}$ (cf.~Eq.~\ref{eq_didt}).

With respect to the LTE assumption, it is unfortunately difficult to estimate in which direction NLTE effects would change the result. On the one hand, NLTE predicts that the same increase in gas temperature will yield a lower increase in intensity than in LTE because the energy is unevenly distributed over the available degree of freedoms, and thus may result in, e.g., ionization imbalances instead of increased radiation. On the other hand, the numerical simulations and experiments by, e.g., \citet{rammacher+ulmschneider1992}, \citet{carlsson+stein1997}, or WE04 yield that the basic process related to propagating waves in the solar chromosphere is a compression of material to hot shock fronts, whose increased temperature leads to enhanced emission. How much the conditions in the relevant small-scale shock fronts and their immediate surroundings then deviate from LTE is not clear at once because there are effects present that can either decrease or increase the source function relative to the Planck function \citep{rammacher+ulmschneider1992,carlsson+stein1997}. It is, however, expected that the LTE approach underestimates the true temperature fluctuations, but by which amount cannot be derived straightforwardly.

The fact that the observed and theoretical curves of temperature fluctuations in the {\em right panel} of Fig.~\ref{fig_rmst_comp} can be roughly matched with a {\em constant} coefficient implies that despite all the simplifying assumptions both the attributed geometrical formation heights and the temperature rms cannot be completely off. Neglecting NLTE effects should have a significant impact above a height of about 400\,km \citep[e.g.,][]{rammacher+ulmschneider1992}, but should presumably introduce an ever increasing deviation with height that is not seen as such. We plan to eventually extend the analysis of the data to a proper NLTE investigation using the inversion code of \citet{socas-navarro+etal2000} in the future, but propose to provide the data\footnote{We remark that both TIP (contact mcv''at''iac.es) and POLIS (contact cbeck''at''iac.es) have an open data policy but for a few exceptions.} used in the present study to anybody interested in improving the diagnostics of the analysis by including NLTE effects.
\paragraph{Mask of magnetic locations:}
The increased skewness at intermediate wavelengths/heights (150\,--\,300\,km) implies more high-intensity events at these layers in the presence of photospheric magnetic fields. We note that for the derivation of the polarisation masks defining field-free and magnetic locations we have only used the circular polarisation signal, because the Stokes $Q$ and $U$ signals are usually one order of magnitude smaller in QS. The magnetic locations thus neither correspond to horizontal nor vertical magnetic fields, but any orientation between these extremes that still produces a measurable circular polarisation signal above the detection limit of the observations (down to a few $10^{-4}$ of $I_c$ for observation No.~3). This covers then network regions, isolated ``collapsed'' structures \citep[i.e., evacuated and concentrated magnetic flux, cf.~][]{stenflo2010}, their immediate surroundings, and also all types of transient polarisation signals above the threshold used. We therefore cannot directly identify which kind of solar magnetic structure is related to increased intensity in the line wing in the statistic sample, but a study of individual events should allow one to pinpoint the corresponding magnetic features.
\subsection{Semi-empirical \emph{vs.}~dynamic chromospheric models}
For the discussion on the validity of semi-empirical \citep[e.g., HSRA, VAL81,
FAL93,][]{fontenla+etal2006,avrett+loeser2008} {\em vs.}~dynamic chromospheric temperature 
models \citep[e.g.,][WE04]{carlsson+stein1997,gudiksen+etal2011} our observations add that 
both \ion{Ca}{ii} lines used here show large rms intensity fluctuations (up to
30\,\%) and a very dynamic behaviour, as also found by \citet{cram+dame1983}
for \ion{Ca}{ii}\,H, and \citet{cauzzi+etal2008} or \citet{vecchio+etal2008}
for the \ion{Ca}{ii}\,IR line at 854.2\,nm. This favors dynamical models of the chromosphere more than the semi-empirical approaches that match average spectra without or with low spatial and temporal resolution.  \citet{avrett2007} states that temperature variations of 400\,K cause an intensity variation in excess of ``{ \em the observed intensity variations at chromospheric wavelengths}''. A similar line of argumentation is used in \citet{kalkofen2012}, going back to a quote from \citet{carlsson+etal1997} that ``{\em all chromospheric lines show emission above the continuum everywhere, all the time}''. There is, however, one big caveat in these claims: the \ion{Ca}{ii} lines were {\em not considered} in the latter works, only the lines and continua observed with SUMER/SOHO and/or atlas spectra. Indeed, the observations of reversal-free Ca profiles challenge a chromosphere which is permanently hot everywhere \citep{rezaei+etal2008}.

The discrepancy between the dynamical and stationary view of the solar
chromosphere here does neither depend on the way of evaluating the data nor
the assumption of LTE or NLTE in the analysis. It is a clear fact that the
observed spectra of \ion{Ca}{ii} lines with their complete change from
(strong) emission to reversal-free absorption profiles contradict the SUMER
observations with permanent and ubiquitous emission with only a minor
variation. As already suggested in BE08, this contradictory behaviour of all
these so-called ``chromospheric'' spectral lines could be reconciled by a
single cause, namely, if the SUMER UV and the \ion{Ca}{ii} lines do not form at the
same height in the solar atmosphere. \citet{carlsson+stein1997} found in their
dynamical NLTE simulations a formation height of \ion{Ca}{ii}\,H of about
1\,Mm, which is about 1\,Mm lower than in static 1D models (e.g., Fig.~1 in
VAL81). The variation in the formation height of the \ion{Ca}{ii}\,H line core is significantly larger than that of photospheric lines. While the former suffers from a spatially and temporally (strongly) corrugated landscape in the source function, the latter benefit from a rather smooth variation of the formation height. Hence, it seems highly desirable to check the formation heights of the UV lines and continua observed with SUMER in dynamical atmosphere models or numerical simulations. It is possible that opposite to \ion{Ca}{ii}\,H their formation heights would have to be significantly raised in a dynamical atmosphere, putting them above an existing magnetic canopy that dampens all oscillations.

\citet{moll+etal2012} recently found that the presence of magnetic fields
strongly affects the flow field in the upper photosphere and consequently the
photospheric and chromospheric heating. As a result, it is important to repeat
dynamic models such as those of WE04 and \cite{carlsson+stein1997} with an
inclusion of magnetic fields. Basically all chromospheric simulations up to
today share simplifications regarding the topology or presence of small-scale magnetic fields, scattering, time-dependent hydrogen ionization, as well as the NLTE radiative transfer \citep{carlsson+etal2012}.
\subsection{Energy sources for the chromospheric radiative losses}\label{sec:radlos}
An atmosphere in radiative equilibrium has no temperature rise. Therefore, an
enhanced chromospheric temperature requires some energy transfer to the upper
layers of the solar atmosphere, resulting in emission of the spectral line
cores. This, however, is distinctly different from the enhanced emission one
finds, e.g., in network bright points. There, a partial evacuation of the
atmosphere and the related shift of the optical depth scale cause the excess
emission. In other words, magnetic bright points and faculae are
(quasi-stationary) bright because of the Wilson depression
\citep{keller+etal2004,steiner05,hirzberger+wiehr2005}. The enhanced skewness
on magnetic locations at intermediate wavelengths therefore could result from the spatial variation of the optical depth scale in the presence of magnetic
fields of variable total magnetic flux. The high-intensity events in the line
wing then should show a close spatial or temporal correlation to the total magnetic flux if they are solely caused by optical depth effects. 

Opposite to that, chromospheric transient brightenings such as bright grains
\citep[e.g.,][]{rutten+uitenbroek1991} are believed to result from a temporary energy
deposit. The intensity enhancements in the line wing could be similar to
bright grains, but the source of the energy deposit in that case is unclear. The primary suspect would be the abundant waves in the solar atmosphere. The propagation and refraction of acoustic and magneto-acoustic waves in and around magnetic flux concentrations can generate strong upflows or downflows
\citep[e.g.,][]{kato+etal2011}. Beside that, the swaying motions of magnetic
elements in the convectively unstable photosphere
\citep{steiner+etal1998,rutten_etal2008} result in shock waves propagating
inside and outside magnetic flux concentrations \citep{vigeesh+etal2011}. Such
spontaneous excitations of shock fronts is another alternative mechanism for
an energy deposit in the lower solar atmosphere. 

Several studies show that a conversion between hydrodynamic waves and different types of magneto-acoustic wave modes is possible near the equipartition layer of similar sound and Alfv\'en speed
\citep[cf.][]{cally2007,haberreiter+finsterle2010,nutto+etal2010a,nutto+etal2010b}. The latter authors investigated wave propagation in numerical
magneto-hydrodynamical (MHD) 2D simulations, in which the equipartition layer can be seen to be located near a height of 300\,km close to a magnetic field
concentration \citep[][their Fig.~1]{nutto+etal2010b}. Nevertheless, pure mode
conversion does not deposit energy into the atmosphere and therefore should
not increase the occurrence rate of high-intensity events. Contrary to that,
\citet{davila+chitre1991} discuss resonance absorption of acoustic waves as a
possible candidate for an energy deposit in the presence of a (horizontal)
magnetic canopy. Numerical simulations should be capable to address this
problem, e.g., by extending the studies of
\citet{leenaarts+wedemeyer2005} and \citet{leenaarts+etal2009} from filter
imaging to an investigation of the full line spectrum of the \ion{Ca}{ii}
lines (including the line blends) and by including magnetic fields into the simulations. This should allow one to determine whether the characteristic differences between magnetic and field-free locations in terms of the skewness found in the observations show up as well in spectra synthesized from MHD simulations. 

Magnetic reconnection or flux cancellation \citep[e.g.,][]{bellot+beck2005,beck+etal2005c} are alternative options for releasing thermal energy in only certain layers of the atmosphere. Such events are more likely to occur in the moat of sunspots
or emerging flux regions where an extensive rearrangement of the magnetic
field lines happens \citep[e.g.,][]{pariat_etal04}. These energetic events
occur below and above the temperature minimum
\citep[e.g.,][]{tziotziou+etal2005,yurchyshyn+etal2010,watanabe+etal2011}.
Even if such high-energy events are not common in QS, granular-scale flux cancellation happens more frequently \citep{reza_etal2007,kubo+etal2010}. The necessary amount of ``new'' magnetic flux for a repeated occurrence of such events can be provided by small-scale flux emergence in the QS \citep[e.g.,][and references therein]{gomory+etal2010} that can effect an energy transport from below the photosphere to the chromosphere. Part of the newly-emerged emergent flux can reach the lower or upper chromosphere \citep{gonzalez+bellot2009}. The two time-series used in the present study (cf.~Fig.~\ref{overview_2009} and BE08) show several cases of transient polarisation signals that could indicate flux emergence events. This allows one a case study of individual brightening events in the line wing on a possible relation to co-spatial or close-by flux emergence, but is beyond the scope of the present study.
\section{Conclusions\label{concl}}
The intensity distributions of the chromospheric lines of \ion{Ca}{ii}\,H and
the \ion{Ca}{ii}\,IR line at 854.2\,nm show a minimum of rms fluctuations for wavelengths forming in the low photosphere caused by the inversion of the granulation pattern. Wavelengths forming above the height of minimal rms show a nearly monotonic increase of rms fluctuations towards the line core that indicates propagating acoustic waves with increasing oscillation amplitudes. A conversion of these intensity fluctuations to corresponding brightness temperature variations yields rms values of about 100\,K in the lower photosphere and a few hundred K in the chromosphere, favoring dynamical models of the solar chromosphere. The fluctuations fall still short of those in numerical simulations and would not suffice to lift an atmosphere in radiative equilibrium to the temperature of, e.g., the HSRA model. The positive skewness for most wavelengths indicates a higher occurrence frequency of high-intensity events, presumably the hot shock fronts formed by the steepening acoustic waves. A prominent difference in skewness between magnetic and field-free locations for wavelengths forming in the mid photosphere indicates the existence of a mechanism that operates only in the presence of magnetic fields and enhances the intensity in the line wing. Single-case studies of such events will allow one to identify the exact process and its relation to the structure of the magnetic field.
\begin{acknowledgements}
The VTT is operated by the Kiepenheuer-Institut f\"ur Sonnenphysik (KIS) at the
Spanish Observatorio del Teide of the Instituto de Astrof\'{\i}sica de Canarias (IAC). The POLIS instrument has been a joint development of the High Altitude Observatory (Boulder, USA) and the KIS. C.B.~acknowledges partial support by the Spanish Ministry of Science and Innovation through project AYA2010--18029. We thank Dr. M. Collados, Dr. H. Balthasar, and Prof. Dr. J. Staude for helpful comments on the manuscript.
\end{acknowledgements}
\bibliographystyle{aa}
\bibliography{references_luis_mod}
\begin{appendix}
\section{Average profiles\label{appa}}
\begin{figure}
\centerline{\resizebox{8.8cm}{!}{\includegraphics{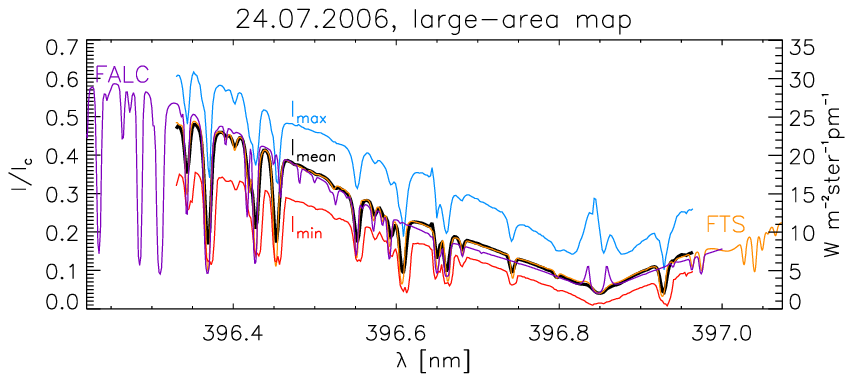}}}
\centerline{\resizebox{8.8cm}{!}{\includegraphics{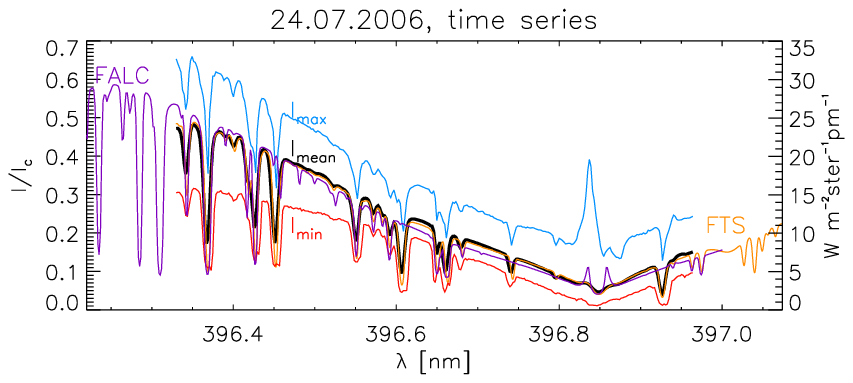}}}
\centerline{\resizebox{8.8cm}{!}{\includegraphics{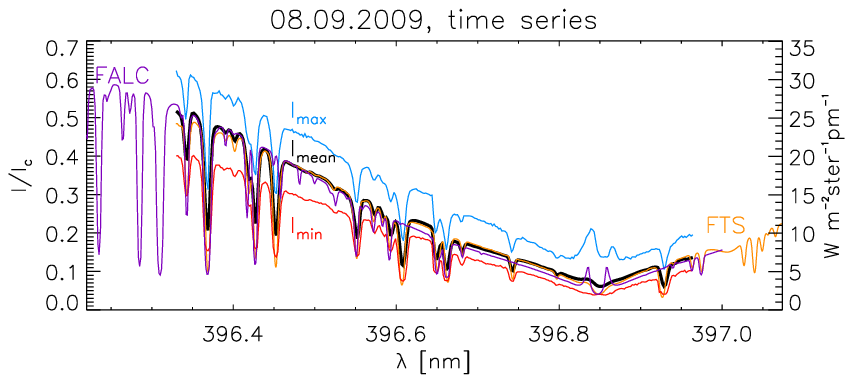}}}
\centerline{\resizebox{8.8cm}{!}{\includegraphics{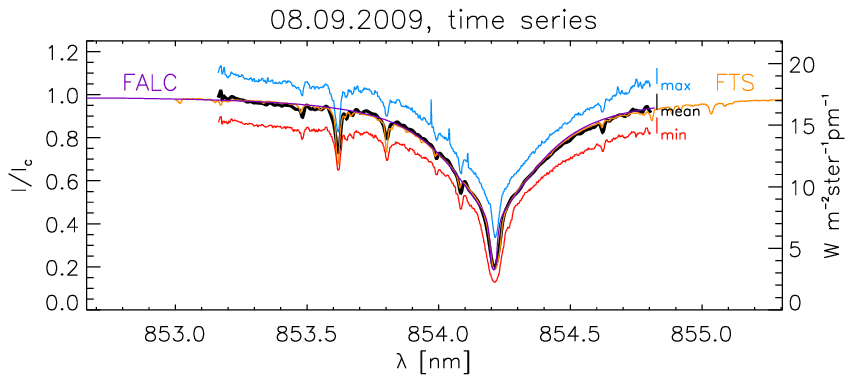}}}
\caption{Average \ion{Ca}{ii}\,H profiles of observations No.~1 ({\em top}), 2 ({\em 3$^{rd}$ row}), and 4 ({\em 2$^{nd}$ row}). {\em Bottom}: average \ion{Ca}{ii}\,IR profile of observation No.~4. Same layout as Fig.~\ref{int_norm}. \label{av_prof1}}
\end{figure}

For completeness, Fig.~\ref{av_prof1} shows the average profiles of the
remaining observations used in this study. The average and minimal intensity
profiles differ only slightly for both the \ion{Ca}{ii}\,IR line at 854.2\,nm and \ion{Ca}{ii}\,H. The maximum observed intensity of \ion{Ca}{ii}\,H varies
significantly from observation to observation opposite to the minimum intensity, implying that extreme excursions towards higher intensity happen more frequently than those towards lower intensity.
\section{Full statistics at selected wavelengths\label{appc}}
Table \ref{tab_stat} lists the complete statistics at three (four) wavelengths
in the \ion{Ca}{ii}\,H (\ion{Ca}{ii}\,IR) line. The skewness of the magnetic
locations is up to three to five times larger than that of the field-free locations for the intermediate wavelengths (396.632\,nm and 854.131\,nm).
\begin{table}
\caption{Intensity statistics at selected wavelengths. All values but the dimensionless skewness are relative to the continuum intensity $I_c$.\label{tab_stat}}
\begin{tabular}{c|c|ccccc}
 $\lambda$ [nm]& region & $I_{\rm min}$ & $I_{\rm av}$ & $I_{\rm max}$ & $\sigma$ & skewness \cr\hline
 396.486 & full FOV & 0.274 & 0.366 & 0.464 & 0.022  &  0.044 \cr
         &field-free& 0.274 & 0.365 & 0.450 & 0.022  &  -0.063\cr
         &magnetic  & 0.288 & 0.373 & 0.464 & 0.024  &   0.282 \cr\hline
396.632 & full FOV   &  0.183 &  0.228 &   0.332  &   0.014 &  0.792\cr
 &field-free& 0.183 &  0.227 &   0.295  &   0.012 &  0.258 \cr
 &magnetic  & 0.190 &  0.236 &   0.332  &   0.018 &  1.221 \cr\hline
 396.847    & full FOV &   0.026&  0.053 &   0.204 &     0.013&  1.854\cr
&  field-free &  0.026 & 0.052  &    0.194&     0.011&  1.602\cr
 &magnetic   &  0.031 & 0.063  &  0.188  &     0.018&  1.563\cr\hline\hline
 853.245 & full FOV &  0.864  & 0.981  &  1.087 & 0.029&  0.002  \cr
& field-free&         0.867  & 0.980  &  1.085 & 0.029&  -0.014\cr
& magnetic  &         0.880  & 0.980  &  1.080 & 0.028&  -0.051\cr\hline
854.131  & full FOV&  0.473  & 0.540  &  0.653 & 0.019&   0.605\cr
 & field-free&        0.473  & 0.539  &  0.616 & 0.017&  0.289\cr
  & magnetic  &       0.482  & 0.553  &  0.653 & 0.024&  0.743\cr\hline
854.180 & full FOV &  0.256  & 0.403  & 0.538 &  0.033&  -0.182\cr
  & field-free&       0.264  & 0.401  & 0.531 & 0.032 &  -0.225 \cr
& magnetic  &         0.268  & 0.410  & 0.535 &  0.034&  -0.036\cr\hline
854.213  & full FOV & 0.128  & 0.189  & 0.338 &0.023  &  0.989\cr
  & field-free&       0.128  & 0.188  & 0.324& 0.021  &  0.744\cr
 & magnetic  &        0.136  & 0.203  & 0.337& 0.030  &  0.931\cr   
\end{tabular}
\end{table}
\section{Intensity response function for the \ion{Ca}{ii}\,IR line at 854.2\,nm\label{appb}}
\begin{figure}
\centerline{\resizebox{8.8cm}{!}{\includegraphics{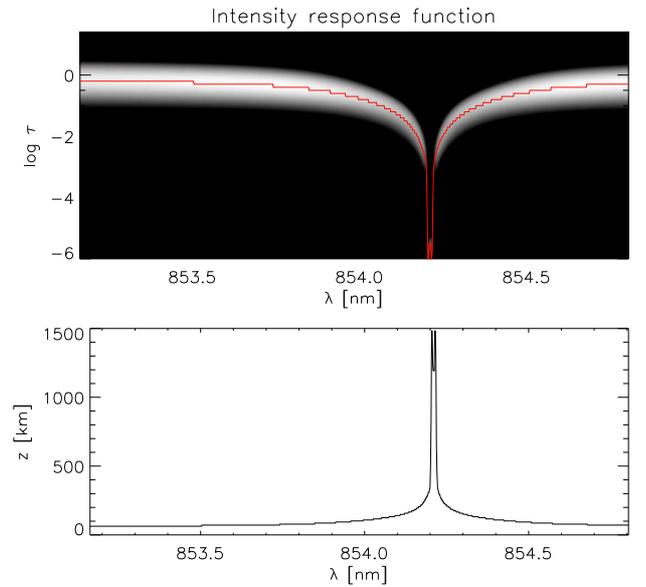}}}
\caption{Conversion from wavelength to geometrical height for the
  \ion{Ca}{ii}\,IR at line 854.2\,nm. {\em Top}: intensity response
  function. The {\em red line} denotes the centre of the response function at each wavelength. {\em Bottom}: geometrical height corresponding to the centre of the response function.  \label{cair_response}}
\end{figure}
To determine the formation height of a given wavelength in the \ion{Ca}{ii}
lines, we synthesized the spectral lines with the SIR code
\citep{cobo+toroiniesta1992} in LTE using a modified version of the HSRA model
\citep{gingerich+etal1971} without chromospheric temperature rise
(cf.~BE09). Using the original HSRA model yielded comparable results (see
Fig.~\ref{err_est} for the case of \ion{Ca}{ii}\,H). The two atmosphere models
are fully identical up to $\log \tau = -4$. We added a temperature
perturbation of 1 K to each of the 75 points in optical depth one after the
other and synthesized the corresponding spectra. The difference to the
unperturbed profile $\Delta I(\lambda, \tau) = I(\lambda,T+\Delta T(\tau)) -
I(\lambda, T)$  yields the intensity response function with wavelength for
perturbations at a given optical depth ({\em upper panel} of
Fig.~\ref{cair_response}). For each $\lambda$, we fitted a Gaussian to $\Delta
I(\lambda,\tau)$, where the centre of the Gaussian yields the optical depth
attributed to that wavelength, $\tau(\lambda)$. With the
tabulated values of $\tau$ and $z$ in the original HSRA model, one can then obtain the corresponding geometrical height, $z(\lambda)$ ({\em lower panel} of Fig.~\ref{cair_response}). The respective curve for \ion{Ca}{ii}\,H was derived fully analogously (see, e.g., BE09). The attributed formation heights from the intensity response function match well those derived from phase differences (BE08, BE09, their Fig.~1). 
\section{Conversion between intensity and brightness temperature fluctuations using the Planck curve\label{appd}}
\begin{figure}
\resizebox{8.8cm}{!}{\includegraphics{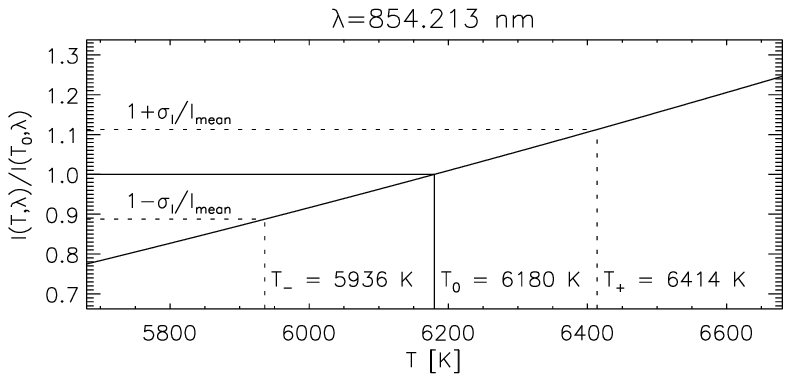}}\\
\resizebox{8.8cm}{!}{\includegraphics{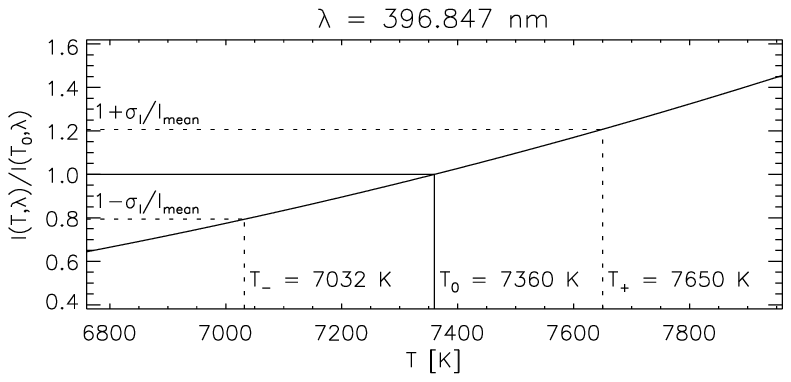}}
\caption{Calculation of brightness temperature variation corresponding to intensity fluctuations of $\pm \sigma_{I}/I_{\rm mean}$. {\em Top}: for $\lambda=854.213$\,nm in the core of the \ion{Ca}{ii}\,IR line.  {\em Bottom}: for $\lambda=396.847$\,nm in the core of the \ion{Ca}{ii}\,H line.\label{conv_i_t}} 
\end{figure}
For the conversion from relative intensity fluctuations to brightness temperature fluctuations, one needs first to attribute a characteristic temperature to the average intensity at each wavelength. To that extent, we used the results of the previous section. The centre of the intensity response function at each wavelength  $\lambda$ yields the optical depth $\tau(\lambda)$ corresponding to that wavelength (e.g., {\em third column} of Table \ref{tab_rms}). The reference atmosphere model then provides the temperature $T_0(\tau)$ at that optical depth, where we assume that the average intensity $I_{\rm mean}(\lambda)$ at the wavelength is directly related to the average temperature $T_0(\tau(\lambda))$ . In the LTE assumption, this relation between intensity and temperature is given by:
\begin{eqnarray}
I_{\rm mean}(\lambda, T_0) = \frac{2 \pi h c^2}{\lambda^5}\cdot \frac{1}{\exp^{\left(\frac{hc}{\lambda k T_0}\right)}-1} \equiv I_0(\lambda, T_0)\,.\label{eq_planck1}
\end{eqnarray}
For a given intensity variation, e.g., by the rms fluctuation $\sigma_I$, one obtains that
\begin{eqnarray}
\frac{I_{\rm mean} \pm \sigma_I}{I_{\rm mean}}  (\lambda)= \frac{I(\lambda, T_0 \pm \Delta T)}{I_0(\lambda, T_0)}\;. \label{eq_planck}
\end{eqnarray}
For $\frac{hc}{\lambda k T_0} \gg 1$, one obtains
\begin{eqnarray}
\frac{\Delta I}{I_0} \propto \frac{\Delta T}{T_0^2}\, =  \frac{\Delta T}{T_0} \cdot \frac{1}{T_0}. \label{eq_didt}
\end{eqnarray}
\begin{table}
\caption{Parameters and temperature fluctuations at selected wavelengths in \ion{Ca}{ii}\,H. \label{tab_rms}}
\begin{tabular}{c|ccccccc}
$\lambda$  & $\frac{\sigma_I}{I_{\rm mean}} (\lambda)$& log $\tau(\lambda)$ &  $z(\tau)$ & $T_0(\tau)$ & $T_-$ & $T_+$  & $\sigma_T$\cr 
nm & & & km & K & K & K& K\cr\hline\hline
396.338&    0.066&    -0.2&    63&       6035&      5966&      6100&    67\cr
396.416&    0.063&    -0.3&      71&       5890&      5830&      5948&   59\cr
 396.492&    0.061&    -0.4&      79& 5765& 5708& 5820&  56\cr
396.494&    0.061&    -0.5&      86&   5650& 5596&  5702&      53\cr
 396.638&    0.061&  -1.0&  125&   5160&      5118&      5200&      41\cr
  396.650& 0.105&  -1.1&  133&       5080&      5006&      5148&      71\cr
 396.693&    0.074&     -1.3&  149&   4950&      4904&      4994&      45\cr
 396.704&    0.076&     -1.4&      158& 4895&      4848&      4940&      46\cr
 396.773&     0.105&     -2.1&      229& 4630&      4568&      4688&      60\cr
 396.778&     0.109&     -2.2&      241& 4600&      4536&      4658&      61\cr
396.784&     0.113&     -2.3&      253& 4575&      4510&      4636&      63\cr
 396.786&     0.114&     -2.4&      266&4550&      4484&      4610&      63\cr
 396.798&     0.124&     -2.6&      294& 4490&      4420&      4554&      67\cr
 396.802&     0.128&     -2.7&      309& 4460&      4388&      4526&      69\cr
 396.804&     0.131&     -2.8&      327& 4430&      4358&      4496&      69\cr
 396.808&     0.135&     -2.9&      344&  4405&      4330&      4472&      71\cr
 396.810&     0.137&     -3.0&      362&  4380&      4306&      4448&      71\cr
 396.812&     0.141&     -3.1&      380& 4355&      4280&      4424&      72\cr
 396.816&     0.150&     -3.2&      399&  4330&      4250&      4402&      76\cr
 396.817&     0.156&     -3.3&      420&4305&      4224&      4378&      77\cr
 396.819&     0.162&     -3.4&      441& 4280&      4196&      4354&      79\cr
 396.821&     0.170&     -3.5&      463& 4250&      4164&      4326&      81\cr
396.823&     0.179&     -3.6&      489&  4225&      4136&      4304&      84\cr
 396.825&     0.192&     -3.7&   515&4205&      4112&      4286&      87\cr
 396.827&     0.206&     -3.8&      540 & 4190&   4090&      4276&      93\cr
396.829&     0.221&     -4.0&      595& 4170&      4064&      4260&      98\cr
396.831&     0.239&     -4.1&      624  & 4200&      4084&      4298&   107\cr
396.833&     0.256&     -4.2&      655& 4280&      4150&      4390&   120\cr
 396.835&     0.273&     -4.4&      721& 4530&  4372&      4660&      144\cr
396.837&     0.287& -4.6&      791&   4790&      4604&      4944&      170\cr
 396.839&     0.294&  -4.8& 873&       5040&      4830&      5214&    192\cr
 396.841&     0.295&     -6.0&      1482&       7360&      6924&  7730& 403\cr
\end{tabular}
\end{table}
Because of the non-linearity of Eq.~(\ref{eq_planck}), $\Delta T$ is different for adding and subtracting the rms fluctuation from $I_{\rm mean}$. We therefore used the average value of $\Delta T$ for the two signs as the temperature variation that corresponds to the rms intensity fluctuation. To obtain the brightness temperature difference corresponding to the increase/decrease of $I$ by one standard deviation $\sigma_I$, the variation of the emergent intensity around the average value $T_0$ is calculated with Eq.~(\ref{eq_planck1}). The relative change of $I/I_0(\lambda, T_0)$ is displayed in Fig.~\ref{conv_i_t} for two wavelengths in the line core of \ion{Ca}{ii}\,H ({\em bottom panel} of Fig.~\ref{conv_i_t}) and  \ion{Ca}{ii}\,IR ({\em top panel} of Fig.~\ref{conv_i_t}). The temperatures corresponding to 1$\pm\sigma_I/I_{\rm mean}(\lambda)$ are then read off from the intersections with the curve of relative intensity variation. Using the modified HSRA model reduces the temperature rms (Fig.~\ref{err_est}) because of the dependence on $T^{-1}$ in Eq.~(\ref{eq_didt}). We used the original HSRA as temperature reference because the majority of the solar atmosphere should presumably be closer to the original than the modified HSRA model. Table \ref{tab_rms} lists the relevant parameters and resulting temperature values for several wavelengths in the \ion{Ca}{ii}\,H line, sorted to have monotonically increasing wavelength and formation heights. We point out that the values of $T_0$ and $z$ were taken from the original HSRA model and were not derived from the observed spectra themselves. 
\end{appendix}
\end{document}